\documentclass[preprint,11pt]{elsarticle}

\usepackage{amssymb}
\usepackage{comment}
\usepackage{subcaption}
\usepackage{amsmath}
\usepackage{tabularray}
\usepackage{setspace}
\usepackage{graphicx}
\usepackage[T1]{fontenc}
\usepackage{booktabs}
\usepackage{tabularx}
\usepackage{siunitx}
\usepackage{caption}
\usepackage{xspace}
\usepackage{lineno}
\usepackage{array}

\newcolumntype{M}[1]{>{\centering\arraybackslash}m{#1}}
\newcolumntype{N}{@{}m{0pt}@{}}

\usepackage{tikz}
\usetikzlibrary{shapes}
\usepackage{xcolor}

\definecolor{myBlue}{rgb}{0.1176, 0.5647, 1.0000}
\definecolor{myRed}{rgb}{0.635,0.078,0.184}
\definecolor{myBlack}{rgb}{0,0,0}
\definecolor{myGreen}{rgb}{0.4660, 0.6740, 0.1880}
\definecolor{myYellow}{rgb}{0.929, 0.694, 0.125}
\definecolor{myPurple}{rgb}{0.3686, 0, 0.6392}
\definecolor{myCyan}{rgb}{0	0.7, 0.741}
\definecolor{myBurd}{rgb}{0.25, 0.058,0}
\definecolor{myOrange}{rgb}{1.0000, 0.4980, 0.3137}

\newcommand{\solidGreen}{\raisebox{2pt}{\tikz{\draw[myGreen,solid,line width=1.5pt](0,0) -- (3mm,0);}}}
\newcommand{\solidBlue}{\raisebox{2pt}{\tikz{\draw[myBlue,solid,line width=1.5pt](0,0) -- (3mm,0);}}}

\newcommand{\solidCyan}{\raisebox{2pt}{\tikz{\draw[myCyan,solid,line width=1.5pt](0,0) -- (3mm,0);}}}
\newcommand{\solidRed}{\raisebox{2pt}{\tikz{\draw[myRed,solid,line width=1.5pt](0,0) -- (3mm,0);}}}
\newcommand{\solidYellow}{\raisebox{2pt}{\tikz{\draw[myYellow,solid,line width=1.5pt](0,0) -- (3mm,0);}}}
\newcommand{\solidPurple}{\raisebox{2pt}{\tikz{\draw[myPurple,solid,line width=1.5pt](0,0) -- (3mm,0);}}}
\newcommand{\solidOrange}{\raisebox{2pt}{\tikz{\draw[myOrange,solid,line width=1.5pt](0,0) -- (3mm,0);}}}

\newcommand{\striangle}{\raisebox{0.5pt}{\tikz{\draw [myBlue,line width=0.3mm,scale=0.07,rotate=270] (0,0) -- (60:3) -- (3,0) -- cycle;}}}

\newcommand{\scircle}[2][myRed,fill=none]{\tikz[baseline=-0.5ex]\draw[#1,radius=#2] (0,0) circle ;}%

\newcommand{\scirclee}[2][myRed,fill=red]{\tikz[baseline=-0.5ex]\draw[#1,radius=#2] (0,0) circle ;}%

\newcommand{\dashLine}{\raisebox{2pt}{\tikz{\draw[black,dashed,line width=0.9pt](0,0) -- (5mm,0);}}}
\newcommand{\dashLineRed}{\raisebox{2pt}{\tikz{\draw[myRed,dashed,line width=0.9pt](0,0) -- (5mm,0);}}}

\newcommand{\solidLine}{\raisebox{2pt}{\tikz{\draw[black,solid,line width=0.9pt](0,0) -- (5mm,0);}}}

\journal{Renewable   Energy}

\begin{document}

\begin{frontmatter}



% \title{Comparison of the mean wake deflection and meandering: large eddy simulation vs. dynamic wake meandering}
%TC:ignore
\title{Validation of the dynamic wake meandering model against large eddy simulation for horizontal and vertical steering of wind turbine wakes} 


\author[inst1]{Irene Rivera-Arreba\fnref{fn1}}

\affiliation[inst1]{organization={Department of Marine Technology, Norwegian University of Science and Technology},%Department and Organization
            %addressline={Jonsvannsveien 82}, 
            city={Trondheim},
            postcode={7050}, 
            country={Norway}}
            
\author[inst2]{Zhaobin Li\fnref{fn1}}
\author[inst2]{Xiaolei Yang\corref{cor1}}
\ead{xyang@imech.ac.cn}

\author[inst1]{Erin E. Bachynski-Poli\'c \corref{cor1}}
\ead{erin.bachynski@ntnu.no}
\cortext[cor1]{Corresponding author}
\fntext[fn1]{These two authors contribute equally.}

\affiliation[inst2]{organization={Laboratory of Nonlinear Mechanics, Institute of Mechanics},%Department and Organization
            addressline={Chinese Academy of Sciences}, 
            city={Beijing },
            postcode={100190}, 
            state={PR},
            country={China}}
\begin{abstract}

This work focuses on the validation of the dynamic wake meandering (DWM) model against large eddy simulation (LES). The wake deficit, mean deflection, and meandering under different wind turbine misalignment angles in yaw and tilt, for the IEA 15MW wind turbine, for two turbulent inflows with different shear and turbulence intensities are compared. Simulation results indicate that the DWM model as implemented in FAST.Farm shows very good agreement with the LES (VFS-Wind) data when predicting the time-averaged horizontal and vertical wake, especially at $x\geq 6D$ and for cases with positive tilt angles ($\geq 6^\circ$). The wake dynamics captured by the DWM model include the large-eddy-induced wake meandering at low Strouhal number ($St < 0.1$). Additionally, the wake oscillation induced by the shear layer at $St \approx 0.27$ is captured only by LES. The mean and standard deviation of the wake deflection, as computed by the DWM, are sensitive to the size of the polar grid used to calculate the spatial-averaged velocity with which the wake planes meander. The power output of a turbine in the wake of a wind turbine in free-wind deflected by a yaw angle $\gamma = 30^{\circ}$ is almost doubled compared to the fully-waked condition. 

\end{abstract}

%TC:endignore
\begin{comment}
\begin{graphicalabstract}
\includegraphics{grabs}
\end{graphicalabstract}

\begin{highlights}
\item Research highlight 1
\item Research highlight 2
\end{highlights}
\end{comment}

\end{frontmatter}

\section{Introduction}
\label{sec:intro}

The need to understand wake effects as a result of the interaction between wind turbines operating in the atmospheric boundary layer is of increasing importance, since they affect the overall performance of a wind farm. Wake effects they imply an increase in power losses \cite{barthelmie2010}. Even though these losses depend on several parameters such as turbulence intensity, turbine spacing, and atmospheric stability \cite{stevens2017}, measurements of a wind turbine completely in wake conditions show up to 40\% losses compared to a wind turbine operating in free flow \cite{barthelmie2010,simley2020}. Furthermore, wake effects cause an increase of fatigue loads for downstream wind turbines, especially for wind speeds below rated and with high turbulence intensity levels. 

The main wake effects are described as deficit and meandering. Additionally, the wake may have a mean vertical or horizontal deflection. 
For onshore and offshore bottom-fixed wind turbines, the horizontal deflection of the wake is a consequence of the misalignment between the main direction of the incoming wind field and the rotor; the vertical deflection is a consequence of the built-in shaft tilt (referred to as \textit{tilt} in the following) and vertical shear. For floating wind turbines, the platform pitch angle also contributes to the effective tilt angle, and thus to the vertical deflection~\cite{wise2019,nanos2021,doubrawa2021}. Both the yaw misalignment and the tilted rotor, combined with the variation in the projected area of the rotor itself, affect the overall performance of the wind farm. 

To reduce the negative effects of the wakes on the downstream wind turbines in a wind farm, several turbine control strategies have been developed and evaluated. A well-studied group of control strategies consists of wake steering, which changes the direction of wakes by deliberately misaligning the rotor and the wind. These strategies include yaw-based \cite{bastankhah2016,fleming2014,howland2016,archer2019} and tilt-based controls \cite{fleming2014,aitken2014a}, %, and torque and pitch control \cite{annoni2015}. 
which deflect the wake in the lateral and vertical directions, respectively. Understanding and predicting how these techniques impact the wake is essential for their implementation in utility-scale wind farms. 

The yaw-based wake control strategy has been a research focus of the wind energy and fluid mechanics community. Howland et al. \cite{howland2016} studied the wake deflection behind a yawed rotor by means of wind tunnel experiments and large eddy simulation (LES). In their study, the wind turbine was represented as a porous disk in uniform inflow. Bastankhah and Porté-Agel \cite{bastankhah2016} also performed wind tunnel measurements, where they studied the characteristics of the wake of a yawed wind turbine by particle image velocimetry. They developed an analytical model to estimate the wake deflection and the far-wake velocity distribution for yawed turbines. Qian and Ishihara \cite{qian2018} also proposed a new analytical wake model for a yawed turbine dependent on the thrust and turbulence intensity. Recently, the success of yaw control has been justified by a multi-month field measurement of a utility-scale wind farm, showing an evident increase of energy production \cite{howland2022collective}. 

%Fleming et al. \cite{fleming2014} and Miao et al. \cite{miao2016} both used CFD to study the effect of the yaw angle of an upstream wind turbine on the performance of two aligned turbines. Both studies showed that the total power output increased if the upstream turbine was positively yawed, whereas if the yaw angle of the upstream turbine was negative, the total power production decreased.\\

The number of studies on the impact of vertical wake steering %on the wake characteristics and on the performance of the wind turbines themselves 
is small compared to the numerous studies on horizontal wake steering, since implementing vertical wake steering is challenging for current bottom-fixed wind turbine designs. Research indicates that vertical steering of the wake also increases the power available to wind turbines downstream. Interestingly, the increase differs for cases with an upward and a downward deflection. In \cite{fleming2014a}, Fleming et al. studied the impact of tilting the rotor on two wind turbines inline, whereas Annoni et al. \cite{annoni2017} did so for three turbines in a row. Both studies reported significant increase in the power output due to upward wake deflection. Johlas et al. \cite{johlas2022} performed large eddy simulations of a 15 MW turbine for different rotor tilt angles to study the impact of downward wake steering on the wake geometry, shear and power production. Consistently with previous studies, they \cite{johlas2022} found that due to the wake being tilted away, and replaced by higher wind speeds from above, the power available to a downwind rotor recovered faster. Nanos et al. \cite{nanos2021} investigated the possibility of deflecting the wake behind a floating wind turbine downwards, by means of imposing a tilt angle by differential ballast control. %Xu et al. \cite{xu2023} focused on the impact of rotor tilt angle on the aerodynamic performance of downwind floating wind turbines, by using a free vortex wake model. They found that the uptilt angle influences the aerodynamic performance of the downwind wind turbine, especially close to the blade root.  
Cossu \cite{cossu2020,cossu2021} suggests that a negative tilt angle is not only able to mitigates the wake effect by steering the wake downwards, but has the potential to increase the wind speed behind the wind turbine by creating high speed streaks in the atmospheric boundary layer.

Previous studies on wake steering have are mainly based on LES, or wind-tunnel experiments. The dynamic wake meandering (DWM) model \cite{larsen2008} represents a compromise between accuracy and efficiency, and several works have aimed to validate it against high-fidelity LES, or experimental data. Churchfield et al. \cite{churchfield2015} compared DWM and
LES with field data from the Egmond aan Zee offshore wind farm; however, the input turbulent wind field for the DWM model was based on a stochastic inflow turbulence generator, whereas for the LES model they used SOWFA, a high-fidelity simulator for the wind turbine dynamics and the fluid flow in a wind farm. Jonkman et al. \cite{jonkman2018validation} validated
the DWM model, as implemented in FAST.Farm, a mid-fidelity tool developed at the National Renewable Energy Laboratory, against SOWFA (LES), for a series of yaw misalignment cases up to 15\textdegree, for a single row of three NREL 5MW turbines separated by eight rotor diameters. In this work, they showed a good agreement between the statistical distribution of the horizontal and vertical wake meandering against LES, suggesting the DWM is a very promising compromise between accuracy and efficiency. However, previous studies did not investigate how the DWM performs when considering wake steering at larger yaw angles, for tilt steering in the vertical direction, and for larger wind turbines.

%{\color{red} To add some modelling approaches for predicting the wake deflection/turbulence (unsteadiness) and say that dynamic wake meandering model \cite{larsen2008} is a compromise between the accuracy and efficiency, but it is not known how well the model performs when considering wake steering, especially in the vertical direction.  }

% {\color{red} For this reason, the objective of the current research is to validate the dynamic wake meandering model against large eddy simulations for }

For this reason, the objective of the current research is to estimate the wake deficit, mean deflection and meandering for different yaw and tilt angles by using FAST.Farm that implements the DWM, for the IEA 15MW \cite{Gaertner2020} wind turbine. The outcome is compared to results computed by large eddy simulations by VFS-Wind, and an analysis on the filter which defines the size of the polar grid used to calculate the spatial-averaged velocity that is used to meander the planes in FAST.Farm, namely $C_\text{meand}$, is provided for different conditions. 
The comparison includes twenty different cases, including 2 inflows with different shear and turbulence intensities, each of them for 10 different wind turbine  misalignment angles (4 yaw, 5 tilt and 1 baseline). %The focus regarding the horizontal wake steering is on the study of positively yawed turbines. Regarding the vertical deflection, both up- and downwards wake deflection are investigated. However, since the studied wind turbine is upwind, downward wake deflection is studied for just one rotor tilt angle case. 
%at four yaw angles and five tilt angles different than 0\textdegree, for two turbulent inflows, at 9\,m/s mean wind speed. FAST.Farm has the dynamic wake meandering (DWM) model implemented to account for wake deficit, deflection and meandering, and the wind turbine is modelled in OpenFAST, which uses the Blade/Element Momentum theory to model the blades. In VFS, the wind turbine wake is modelled using LES, and the blades and nacelle are modelled as actuator surfaces \cite{yang2018ASMethod}. 
The focus is on the time-averaged velocity field and wake deficit, the average wake deflection, vertical and horizontal, and the meandering, or standard deviation of the wake center displacement, at 8 rotor diameters ($D$) downstream, which is a relevant spatial interval for turbines in a wind farm. Finally, the effect of the yaw and tilt deflection on the power output at a wind turbine placed $8D$ downstream is investigated in FAST.Farm.

The remainder of the paper is structured as follows. Section~\ref{sec:methodology} describes the two methods compared throughout this work: LES by VFS-Wind and the DWM model as implemented in FAST.Farm. Section~\ref{sec:numerical} gives an overview of the simulation set-up, the wind turbine model, the generation of the wind field and the characteristics of the computational domain in VFS and FAST.Farm. Section~\ref{sec:results}  analyses the time-averaged deficit and wake meandering of FAST.Farm with respect to VFS, and their effects on the power output of a turbine placed downstream. The conclusions on this work are retrieved in Section~\ref{sec:conclusions}.

\section{Methodology}\label{sec:methodology} 

\subsection{Large eddy simulation by VFS-Wind}

We employ the large eddy simulation module of the Virtual Flow Simulator (VFS-Wind \cite{yang2015VFS}) to provide high-fidelity wake simulations. The solver treat the airflow as a Newtonian fluid with a constant density and viscosity. The governing equation is the filtered incompressible Navier-Stokes equations, as follows:

\begin{align}
	% \frac{\partial\widetilde{u_i}}{\partial x_i} & =0,\label{eqn:cnt}\\
 %    \frac{\partial\widetilde{u_i}}{\partial t} + \frac{\partial\widetilde{u_i} \widetilde{u_j}}{\partial x_j} & = -\frac{1}{\rho}\frac{\partial \widetilde{p}}{\partial x_i} + \nu \frac{\partial^2 \widetilde{u_i}}{\partial x_j \partial x_j} - \widetilde{\tau_{ij}} + \frac{1}{\rho}f_i
\nabla \cdot \widetilde{\mathbf{u}} & = 0 ,\label{eqn:cnt} \\ 
\frac{\partial \widetilde{\mathbf{u}}}{\partial t}+(\widetilde{\mathbf{u}} \cdot \nabla) \widetilde{\mathbf{u}} & =-\frac{1}{\rho} \nabla \widetilde{p} + \nu \nabla^2 \widetilde{\mathbf{u}} - \nabla \cdot \mathbf{\tau}  + \frac{\mathbf{f}}{\rho},
	 \label{eqn:ns}
\end{align}
where $\mathbf{u} = \{u_x, u_y, u_z\}$ is the velocity vector in the Cartesian coordinates,  $p$ is is the pressure, $\rho$ is the fluid density, $\nu$ is the fluid kinematic viscosity.  $\widetilde{\cdot}$ denotes the spatial filtering, and $\mathbf{\tau}$ is the subgrid-scale stress and is closed with the eddy-viscosity model \citep{Smagorinsky}. $\mathbf{f}$ is a body force term employed to represent the effect of wind turbines on the flow and is computed using a well-validated actuator surface model \citep{yang2018ASMethod} for both the rotor and nacelle, described in section~\ref{sec:windTurbineModel}.

The governing equations are discretized using the finite differencing method on a structured rectangular grid. The spatial discretization uses the second-order central differencing scheme. The temporal integration employs a second-order fractional step scheme.  The nonlinear momentum equation is solved by the Jacobian-free Newton-Krylov approach \citep{knoll2004jacobian}. The Poisson equation, derived from the continuity equation to enforce incompressibility, is solved by the  Generalized Minimal Residual (GMRES) approach \citep{saad1993flexible} with multigrid as a preconditionner. A detailed description of the numerical implementation can be found in  \cite{ge2007numerical}.

The predictive capability of VFS-Wind has been systematically validated against wind tunnel and field experiments \cite{yang2016coherent,hong2020snow}. It has been widely applied to reveal the fluid mechanics of wind turbine wake \cite{foti2021coherent,li2021,li2022onset},  and to investigate complex flow in large-scale wind farms \cite{yang2018large,wang2023statistics}.

\subsection{Dynamic wake meandering model in FAST.Farm}

The basis behind the mid-fidelity dynamic wake meandering (DWM) model is the division of turbulence scales by a \textit{cut-off eddy size} filter: turbulent eddies smaller than this size affect the evolution of the wake deficit, whereas the larger eddies mainly impact wake meandering. This size can be a model parameter, but it is usually taken to be two rotor diameters. The original model as proposed by Larsen et al. \cite{larsen2008} consists of three submodels: the wake deficit, the wake meandering and the added-wake turbulence. The wake deficit evolution is described in the meandering frame of reference, and is modelled by the thin shear-layer approximation of the Reynolds-averaged Navier-Stokes equations under quasi-steady-state conditions in axisymmetric coordinates, in the far-wake region, as:

\begin{equation}\label{eq:deficitEq}
\centering
    U\frac{\partial U}{\partial x}+V_r \frac{\partial U}{\partial r} = \frac{\nu_T}{r}\frac{\partial}{\partial r}\left(r\frac{\partial U}{\partial r}\right),
\end{equation}

\noindent where $U$ is the axial velocity component, $V_r$ the radial velocity component, $r$ the radial coordinate and $\nu_T$ the eddy viscosity. The turbulence closure is modelled by an eddy-viscosity formulation, which assumes that the velocity gradients are higher in the radial direction as compared to those in the axial direction; the pressure term is neglected.

FAST.Farm \cite{jonkman2021} is a multiphysics engineering tool used to predict the power performance and structural loads of wind turbines within a wind farm. This software uses OpenFAST (version 3.4.0 \cite{OF34} in the current work) to solve the aero-hydro-servo-elastic dynamics
of each individual turbine, and is based on the implementation of the DWM to account for wake deficit and meandering. In FAST.Farm, the eddy-viscosity is modelled by the longitudinal distance $x$-dependent filter parameters $F_{amb}$ and $F_{shear}$. These parameters were described and calibrated by Madsen et al. \cite{madsen2010}, and extended by Larsen et al. \cite{larsen2008} and Keck \cite{keck2013PhD}. The eddy-viscosity $\nu_T(x,r)$ dependent on $x$ and the radial position $r$ as implemented in FAST.Farm is:

\begin{equation}
\begin{aligned}
\nu_T(x,r) & =  \nu_{amb}(x)  +  \nu_{shear}(x,r) =\\ 
& = F_{amb}(x) k_{amb}\text{TI}_{amb} V_{x}\frac{D_w}{2} + \\ 
&  + F_{shear}(x) k_{shear}\text{max}\left\{\left(\dfrac{D_{w}}{2}\right)^2\left|\frac{\partial V_x}{\partial r} \right|,\frac{D_{w}}{2}{V_{x,min}} \right\},
\end{aligned}
\end{equation}

\noindent where $k_{amb}$ and $k_{shear}$ are parameters that weight the ambient and shear turbulence influence on the eddy-viscosity, $V_x$ is the time-filtered disk average wind velocity normal to the actuator disk, TI$_{amb}$ is the ambient turbulence intensity at each rotor and $D_{w}$ is the wake diameter. The filter functions $F_{amb}$ and $F_{shear}$ depend on user-specified calibrated parameters. The ones used in this research are based on the work of Doubrawa et al. \cite{doubrawa2018}.

FAST.Farm allows for both a polar and a curl formulation for the wake. The former implies that the wake, which is defined on a polar grid, is axisymmetric. In the latter, the wake is defined on a Cartesian grid and the effect skewed inflow is accounted for by introducing cross-flow velocities. In this work the polar formulation is applied, since at the time the present work was performed, the curled-wake formulation in FAST.Farm was at the development stage. %Since the focus of this research is the skewed inflow, both formulations are used, and compared. 
The radial increment of the radial finite-difference grid is set to 5\,m and the cut-off frequency of the low-pass time-filter $f_c$ for the wake advection, deflection, and meandering model is 0.1\,Hz, based on the work by Branlard et al. \cite{branlard2022}. They define the filter frequency $f_c$ in terms of the time scale $\tau_1$ used in the Øye dynamic inflow model \cite{Oye1991}, as:

\begin{equation}
f_c = \dfrac{2.4}{\tau_1} \quad \quad \text{with} \quad \quad \tau_1 = \dfrac{1.1\cdot D}{2 U_{\infty,z_{HH}}\cdot(1 - 1.3 \text{min}(a_{avg},0.5))},
\end{equation}
where $a_{avg}$ is the time-averaged induction factor and $U_{\infty,z_{HH}}$ the incoming undisturbed wind speed at hub-height $z_{HH}$. The remaining parameters, namely the calibrated parameters related to the near-wake correction, to the eddy-viscosity filter and to the wake diameter calculation, respectively, are set to the default values. A windowed jinc function is used to apply the spatial filter parameter for the wake meandering, defined as $C_\text{meand}$ in FAST.Farm, and chosen on a case-to-case basis, since it directly affects the meandering. This filter, proposed by Larsen et al. \cite{larsen2008}, consists of taking the uniform spatial average where all points within a circle of diameter $2D$ are given equal weight. Based on the windowed jinc function, a value of $C_\text{meand} = 2 $ in FAST.Farm results in a polar-grid diameter of $2D$ and cut-off wave number of 1/$2D$, consistent with the proposed filter of Larsen et al. In other words, $C_\text{meand}$ determines the size of the polar grid used to calculate the spatial-averaged velocity that is used to meander the planes. Therefore its value impacts the amount of wake deflection that occurs. With a larger $C_\text{meand}$, the polar grid is larger, the wake deficit has less influence, and consequently the deflection is smaller. Cheng and Porté-Agel \cite{Cheng2018} proposed a low-pass-filter threshold proportional to the time delay due to downstream advection based on Taylor's diffusion theory, and therefore on the downstream distance $x$. This filter size is $ \displaystyle \dfrac{x}{\bar{u}_{HH} \alpha}$, with $\alpha = 3$ and $\bar{u}_{HH}$ the mean value of the incoming velocity at hub-height. Brugger et al. \cite{Brugger2022} found that, based on field measurements at a utility-scale wind turbine, this filter gave similar results to the one proposed by Larsen et al. up to 6D downstream, but the correlation between the wake meandering and the $v$-component by using this filter improved further downstream. In this work, the value for $C_\text{meand}$ is chosen based on the best fit to the VFS data, and the importance of this parameter selection is discussed.

\section{Numerical set-up}\label{sec:numerical}

\subsection{Simulation set-up}

\begin{figure}[h!]

\centering
\includegraphics[trim={0 0.9cm 0 0cm},clip,width=1\textwidth]{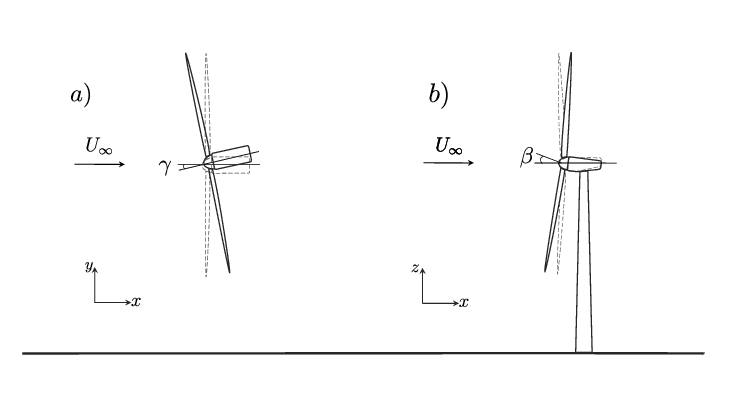}

\centering
\caption{\label{fig:angles}a) Top-view of the yawed rotor with a positive yaw misalignment angle $\gamma$ indicated. b) Side-view of the tilted rotor with the positive tilt angle indicated as $\beta$. $U_{\infty}$ is the incoming undisturbed wind speed. }

\end{figure}

We compare the predictive capability of FAST.Farm and VFS-Wind in 20 different scenarios, namely 10 different yaw/tilt angles (four yaw angles $\gamma$ and five tilt angles $\beta$, and one case with  $\gamma=\beta=0$), and each of them for 2 turbulent inflow conditions. 
%Both FAST.Farm and VFS-Wind simulate twenty cases are employed to study the vertical and horizontal wake deflection due to yaw misalignment of the rotor and to the rotor tilt,: one case without yaw or tilt, plus four yaw angles $\gamma$ and five tilt angles $\beta$ all simulated 
A positive yaw misalignment angle $\gamma$ and a positive tilt angle $\beta$ are depicted in Figure \ref{fig:angles}. Table~\ref{tab:cases} presents the yaw misalignment angle and the tilt angle for every simulated case. For every case in Table~\ref{tab:cases}, the $C_\text{meand}$ in FAST.Farm that yielded the closest mean and standard deviation of the wake mean deflection and standard deviation compared to VFS, measured at $8D$ downstream, is provided in Table \ref{tab:cMeand}. For the cases with yaw misalignment (1a to 4a) the mean value of this filter is 2.0, and for the cases with tilt deflection (1b to 5), 3.2. For the case with no yaw misalignment or tilt deflection angle (case 0), $C_\text{meand}$ is set based on the work of Cheng and Porté-Agel \cite{Cheng2018}, which yields a higher value than the default one proposed in the FAST.Farm manual \cite{jonkman2021}. The effect of using a different filter $C_\text{meand}$ is analysed in section~\ref{sec:Cmeand}.

\begin{table}[h!]
\caption{ \label{tab:cases} Simulation cases for the two scenarios of yaw misalignment (\textit{Yaw}) and tilt deflection (\textit{Tilt}). }
\centering
 \resizebox{\linewidth}{!}{\begin{tabular}{c c| c c c c | c  c c c c }
Case &   0   & 1a        & 2a       & 3a        & 4a               & 1b  & 2b  &  3b   &   4b  & 5  \\
\hline
Yaw  & $\beta$ = 0$^{\circ}$             & $\gamma$ = 10$^{\circ}$    & $\gamma$ = 15$^{\circ}$    & $\gamma$ = 20$^{\circ}$   & $\gamma$ = 30$^{\circ}$  &\color{gray} $\gamma$ = 0$^{\circ}$  &\color{gray} $\gamma$ = 0$^{\circ}$  & \color{gray}$\gamma$ = 0$^{\circ}$  & \color{gray}$\gamma$ = 0$^{\circ}$  &\color{gray} $\gamma$ = 0$^{\circ}$  \\
\hline
Tilt & $\gamma$ = 0$^{\circ}$  & \color{gray}  $\beta$ = 0$^{\circ}$  & \color{gray} $\beta$ = 0$^{\circ}$  &\color{gray} $\beta$ = 0$^{\circ}$  &\color{gray} $\beta$ = 0$^{\circ}$  & $\beta$ = -6$^{\circ}$  & $\beta$ = 6$^{\circ}$  & $\beta$ = 10$^{\circ}$  & $\beta$ = 15$^{\circ}$  & $\beta$ = 20$^{\circ}$       \\
    \hline
    \end{tabular}}
\end{table}

\begin{table}[h!]
\caption{ \label{tab:cMeand} Fitted $C_\text{meand}$ filter values based on the results at $x = 8D$ downstream, for both incoming inflows. }
\centering
 \resizebox{\linewidth}{!}{\begin{tabular}{c c | c c c   c | c c c c c }
Case & 0  & 1a        & 2a       & 3a        & 4a               & 1b  & 2b  &  3b   &   4b  & 5  \\
\hline
  $C_\text{meand}$         & 2.67 & 1.90   & 2.00   & 2.00 & 2.10  & 2.90 & 2.90 & 3.22 & 3.25 & 3.50\\
\hline
    \end{tabular}}
\end{table}
\subsection{Wind turbine model set-up \label{sec:windTurbineModel}}

The reference IEA 15MW wind turbine \cite{Gaertner2020}, with a rotor diameter of 240\,m and 150\,m hub-height (HH), is used. Since the focus of this research is on validating the DWM against LES, the turbine rotor model and operational parameters were defined as similarly as possible in the FAST.Farm  and the VFS models. To this end, a simple control strategy with a fixed rotational speed is employed in both methods. The rotational speed is fixed to $\widetilde{\Omega} = 6.4$\,rpm for the case without yaw and tilt, and it varies with $\cos \theta$ ($\theta$ for yaw or tilt angles), to keep the tip speed ratio $\lambda = \widetilde{\Omega}\cos \theta {D}/(2 U_\infty \cos \theta) $ constant with respect to the rotor normal wind speed. We note that this simple strategy is only to ensure a fair comparison between both models and may not represent the optimal control strategy under yaw and tilt misalignment.      

In VFS, the aerodynamics of the wind turbine were computed using the actuator surface (AS) model proposed by Yang and Sotiropoulos \cite{Yang2019}. This model simplifies the rotor blades as zero-thickness force-carrying surfaces. The aerodynamic forces are computed along the blade using two-dimensional airfoil coefficients, as follows:
\begin{equation}
\mathbf{f} = \mathbf{F}/c  = (\mathbf{L}+\mathbf{D})/c,  
\end{equation}

%\begin{equation}
   % \mathbf{F} = \mathbf{L} + \mathbf{D} 
%\end{equation}
\begin{equation}
\mathbf{L} = \frac{1}{2}c\rho C_\text{L}(\alpha, Re_c)|V_\text{ref}|^2 \mathbf{e_\text{L}}
\end{equation}
\begin{equation}
\mathbf{D} = \frac{1}{2}c\rho C_\text{D}(\alpha, Re_c)|V_\text{ref}|^2 \mathbf{e_\text{D}}, 
\end{equation}

\noindent where $\mathbf{f}$ is the aerodynamic force $\mathbf{F}$ per unit span, with $\mathbf{L}$ and $\mathbf{D}$ the lift and drag, and $C_L(\alpha, Re_c)$ and $C_D(\alpha, Re_c)$ the lift and drag coefficients, depending on the local airfoil and the angle of attack ($\alpha$)  and the Reynolds number ($Re_c$) defined with the chord length ($c$). $V_\text{ref}$ is the relative flow velocity with respect to the rotating blade,  $\textbf{e}_\text{L}$ and $\textbf{e}_\text{D}$ are unit vectors defining the lift and drag directions. 3D effects \citep{du19983Dstall} and tip losses \citep{shen2005tiploss2} are corrected before computing the momentum source in equation \eqref{eqn:ns}.
%

%
%The actuator surface model for the nacelle enforces the non-penetration boundary condition in the nacelle surface's normal direction and applies the frictional force with empirical friction coefficients \citep{yang2018ASMethod}. A smoothed discrete delta function  \citep{yang2009Kernel} is employed to map these forces to the background grid nodes to avoid singularity issues. 

% \color{red}ACTUATOR SURFACE. Introduce here at some point, if possible, the rotor angular velocity $\tilde{\Omega}$:

% \begin{equation}
% \tilde{\Omega} =\frac{2 \tilde{\lambda} U_p}{D},
% \end{equation}

% where $\tilde{\lambda}$ is the tip speed ratio and the projected velocity on the rotor plane $U_p$ is:

% \begin{equation}
% U_p=U_{\infty} \cos \theta, 
% \end{equation}

% where $U_{\infty}$ is the mean longitudinal wind speed component at hub-height and $\theta$ can be the tilt or yaw angle.\\
\color{black}

In OpenFAST, the loads on the rotor are computed based on the blade-element/momentum (BEM) theory by Glauert \cite{Glauert1935}. To reduce the sources of uncertainty when comparing the wake deflection between VFS and FAST.Farm, the wind turbine parameters in OpenFAST are set so that the rotor angular speed $\tilde{\Omega}$ is fixed to the same value as for the actuator surface in VFS. Additionally, the blades in OpenFAST are modelled as rigid.

\begin{figure}[h!]
\centering
    \sbox0{\scircle{3pt}}\sbox1{\striangle}%
    \centering
    \framebox{\usebox0   FAST.Farm, \usebox1 VFS-Wind}
    
    \vspace{0.5cm} 
    
\includegraphics[width=1\textwidth]{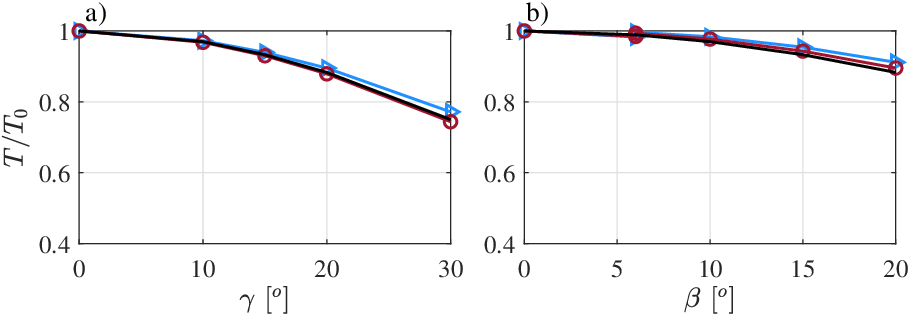}
\caption{ \label{fig:IEAfigure3} Thrust as a function of the yaw ($\gamma$) and tilt ($\beta$) angles, normalized by $T_0$: a) for the yaw misalignment angle cases, and b) for the tilt angle cases. The black solid line represents the $\cos^2 \theta$ relation, with $\theta = \gamma$ or $\beta$.  }
\end{figure}

To verify the similarity of the set-up of the wind turbine in both models, Figure~\ref{fig:IEAfigure3} compares the thrust $T$ of the rotor for the two models as a function of yaw and tilt angles ($\gamma$ and $\beta$). The thrust is defined as normal to the rotor plane for every case shown here. The values are normalized by $T\textsubscript{0}$, which corresponds to the thrust normal to the rotor plane when the yaw misalignment or the tilt angle is zero. Given the control strategy in the current study, the thrust $T$, follows closely the relation $\cos^2 \theta$, with $\theta = \gamma$ or $\beta$, in the two models and for every tilt or yaw angle case, which is consistent with previous findings \cite{Burton2011,Krogstad2011,Bartl2018}. The thrust coefficient $\tilde{C_T}$ and the rotational speed $\tilde{\Omega}$ for zero tilt and yaw angles, are presented in Table~\ref{tab:tilde}. %Figure~\ref{fig:IEAfigure3} compares the thrust $T$ and power $P$ at the rotor for the two models, as a function of yaw and tilt angles ($\gamma$ and $\beta$). The values are normalized by $T\textsubscript{0}$ and $P\textsubscript{0}$, which correspond to the thrust normal to the rotor plane and the power output, respectively, when the yaw misalignment or the tilt angle is zero. Given the control strategy in the current study, $T \propto \cos^2(\gamma)$ and $P \propto \cos^3(\gamma)$, consistently with previous findings \cite{Burton2011,Krogstad2011,Bartl2018}. The black solid line represents the $\cos^2 \theta$ and $\cos^3 \theta$ relationship, for thrust and power, respectively, with $\theta = \gamma$ or $\beta$. The thrust and power coefficients, $\tilde{C}_T$, $\tilde{C}_P$, and the rotational speed $\tilde{\Omega}$ for zero tilt and yaw angles, are presented in Table~\ref{tab:tilde}.
\begin{comment}
Therefore, the thrust and power coefficient independent of the yaw or tilt angle, $\tilde{C}_T$ and $\tilde{C}_P$, read:

\begin{equation}
\begin{aligned}
\tilde{C}_T & =\frac{T}{\frac{1}{2} \rho A U_p^2}=\frac{T}{\frac{1}{2} \rho A\left(U_{\infty} \cos \theta\right)^2}, \\
\tilde{C}_P & =\frac{P}{\frac{1}{2} \rho A U_p^3}=\frac{P}{\frac{1}{2} \rho A\left(U_{\infty} \cos \theta\right)^3},
\end{aligned}
\end{equation}
where $U_{\infty}$ is the undisturbed incoming wind speed, $A$ the rotor area, and $U_p =  U_\infty \cos{\theta}$ is the rotor normal velocity.

$\tilde{C}_T$ and $\tilde{C}_P$, together with $\tilde{\Omega}$, are presented in Table~\ref{tab:tilde}.
\end{comment}

\begin{table}[h!]
\caption{ \label{tab:tilde} Thrust coefficient $\tilde{C}_T$ and rotational speed  
for zero yaw and tilt angles. }
\centering
 \begin{tabular*}{0.2\textwidth}{c c}
$\tilde{C}_T$  &    $\tilde{\Omega}$   \\
 \hline
 0.78    & 6.4\,rpm        \\
    
    \hline
    \end{tabular*}
\end{table}

%{\color{red} NEED to be modified! I suggest to remove Cp from the comparison! }

\subsection{Turbulent inflow wind generation and characteristics}

For every case presented in Table~\ref{tab:cases}, the turbine is subjected to two incoming turbulent wind fields, with different turbulence levels. The incoming wind fields are generated by large eddy simulations using VFS-Wind, with a half-channel configuration to mimic the atmospheric boundary layer flow.  
The precursory LES employed a large computational domain to capture large-scale turbulent structures in the atmospheric boundary layer \cite{wang2016very,liu2019three}, which measures approximately $22.5~\textrm{km} \times 15~\textrm{km} \times 1~\textrm{km}$ in the lateral, streamwise, and vertical directions.  Periodic boundary conditions were applied in both horizontal directions. The flow is controlled to have a constant streamwise flux. The top boundary was free-slip. The lower boundary was a no-slip rough wall modeled with the logarithmic wall functions, i.e., $u/u^* = \displaystyle\frac{1}{\kappa} \ln(z/z_0) $, where $u^{*}$ is the friction velocity, $\kappa$ is the K\'{a}rm\'{a}n constant and $z_0$ is the roughness length \cite{blocken_cfd_2007}.  The roughness length $z_0$ is the control parameter to generate inflows with different vertical shears and turbulence intensities. The streamwise and spanwise grid intervals are  $\Delta x= 20 ~\text{m}$  and $\Delta y = 10$ m, respectively. In the vertical direction, the grid is refined towards the ground and has a grid size of approximately $\Delta z \approx 2$ m in the first layer. The present resolution results in a total number of grid points equal to 254 million. The time step for the simulation is $\Delta t =  1$\,s.  In the precursory simulations, instantaneous velocity fields on a plane perpendicular to the mean flow direction are saved each time step and applied as inflow conditions for the turbine wake simulations in VFS and FAST.Farm. If the resolutions were different, linear interpolations were carried out in space and time. In FAST.Farm, the wind fields generated by LES used as input were linearly interpolated to a time step $\Delta t = 0.1$\,s and $\Delta z = \Delta y = 8$\,m.

%. (\color{red}Include The resolution in y and z is 8.00\,m and the time step is 0.1\,s if Zhaobin does not mention it \color{black}. Zhaobin: I have mentioned the grid resolution employed for the precursor LES simulation in the previous paragraph, but I think you need to mention the one you employed in FF for the wake simulation. ).\\

\begin{figure}[h!]
\centering
    \sbox0{\solidPurple}\sbox1{\solidYellow}%
    \centering
    \framebox{\usebox0   inflow 1  , \usebox1 inflow 2 } 
\centering

    \vspace{0.5cm} 
    
\includegraphics[width=0.98\textwidth]{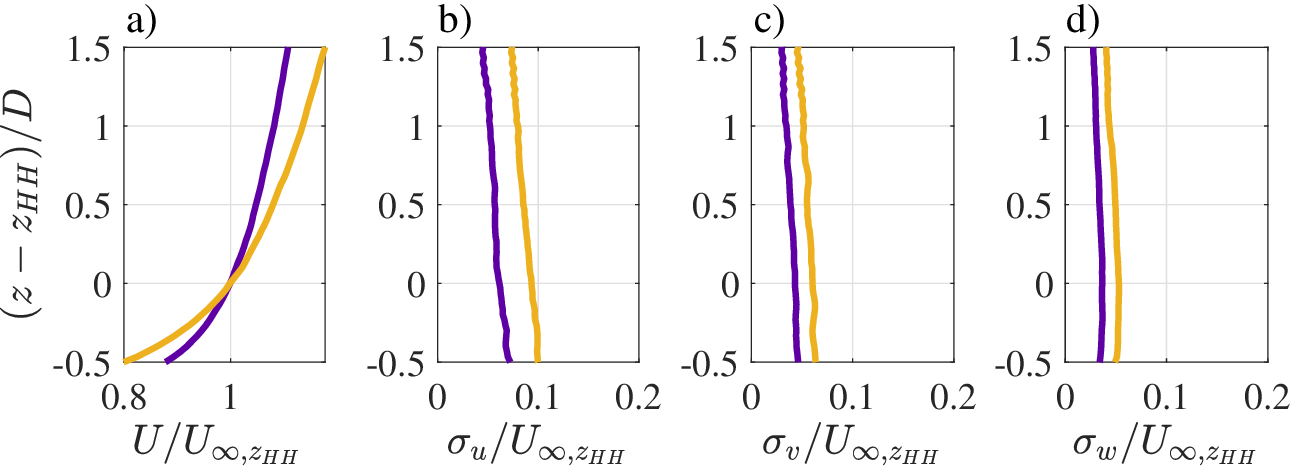}
\caption{ \label{fig:IEAfigure2} Time-averaged turbulent inflow characteristics for the two inflows: a) shear profile of the longitudinal component $U$; b, c, d) standard deviation of the longitudinal, transverse and vertical components, $\sigma_u$, $\sigma_v$ and $\sigma_w$, respectively. All the values are normalized by the undisturbed incoming wind speed at hub-height $U_{\infty,z_{HH}}$. }
\end{figure}

\begin{figure}[h!]
\centering
    \sbox0{\solidPurple}\sbox1{\solidYellow}\sbox2{\dashLineRed}\sbox3{\dashLine}%
    \centering
    \framebox{\shortstack{\usebox0   inflow 1 (I1), \usebox1 inflow 2 (I2)\\
    \usebox2 IEC Kaimal I1, \usebox3 IEC Kaimal I2}} 
\centering

    \vspace{0.5cm} 
    
\includegraphics[width=0.98\textwidth]{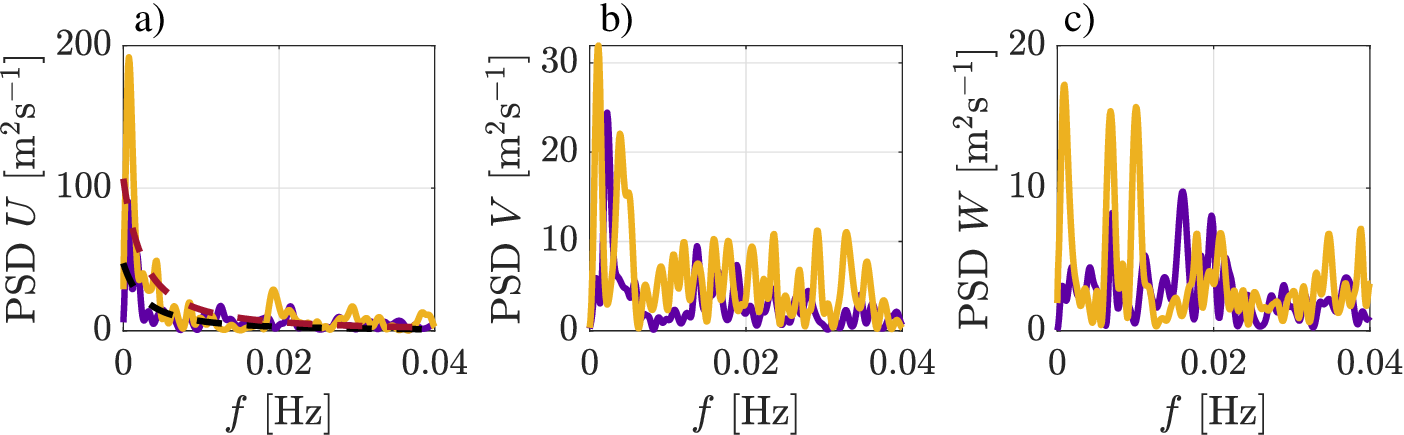}

\caption{ \label{fig:PSDWF} Power spectral density (PSD) of the longitudinal $U$ (a), lateral $V$ (b) and vertical $W$ (c) components of the wind speed at hub-height, for the two inflows. The red and black dashed lines in a) represent the IEC Kaimal spectra for inflow 1 and 2. }
\end{figure}

The shear profile of $U$ along the height $z$ is generally described by the power-law as:

\begin{equation}
U(z)=\bar{u}_{ref}\left(\dfrac{z}{z_{ref}}\right)^\alpha,
\end{equation}

\noindent where $\bar{u}_{ref}$ is the wind speed at a reference height $z_{ref}$ and $\alpha$ is the power-law exponent. The $\alpha$ exponent is a bulk parameter which includes the effect of atmospheric stability and of surface roughness $z_0$ \cite{irwin1979, emeis2014}. Figure~\ref{fig:IEAfigure2} a) presents the shear profile of the longitudinal wind speed component, normalized by the undisturbed incoming wind speed at hub-height $U_{\infty,z_{HH}}$, and computed as the average along the rotor span, i.e. $-D/2<y<D/2$, for every height. The two incoming wind fields differ in turbulence intensity level (TI), which is defined as:

\begin{equation}
    \text{TI} = \dfrac{\sigma_{u}}{U_{\infty,z_{HH}}},
\end{equation}

\noindent where $U_{\infty,z_{HH}}$ is 9\,m/s, for both inflow 1 and inflow 2. $\sigma_u$ is the standard deviation of the longitudinal wind speed component, computed as the average of the standard deviations for points along the rotor span ($-D/2<y<D/2$) for every height. Figure~\ref{fig:IEAfigure2} b, c and d) presents the average of the standard deviation computed across the rotor ($-D/2<y<D/2$) for every height, of the longitudinal, lateral and vertical wind speed components with height z. The relationship between the longitudinal, lateral, and vertical components at hub-height is presented in Table \ref{tab:inflow}, together with the $\alpha$ exponent and the TI of the two turbulent inflows used in this work. The values of $\sigma_v$ and $\sigma_w$ which are only slightly different from the values of $\sigma_v = 0.85 \sigma_u$ and $\sigma_w = 0.60 \sigma_u$ provided in the IEC standards \cite{IEC61400-1}, for neutral atmospheric conditions, and low terrain complexity. Furthermore, the standard deviation is seen to barely vary with height. The power spectral density (PSD) of the three wind speed components for the two inflows is presented in Figure~\ref{fig:PSDWF}, where the decrease in energy as the frequency increases is observed. The dashed black and red lines in Figure~\ref{fig:PSDWF} a), for the $U$-component, represent the IEC Kaimal spectra \cite{IEC61400-1}, for the two inflows with different turbulence levels, respectively. If the Kaimal spectra are compared to the corresponding ones related the conditions in this work, the former underestimate the frequency content at the lowest frequencies range.

\begin{table}[h!]
\caption{ \label{tab:inflow} Shear exponent $\alpha$, turbulence intensity TI and standard deviation of $v$ and $w$, $\sigma_v$ and $\sigma_w$, of the two turbulent inflow wind fields, inflow 1 (I1) and inflow 2 (I2). }
\centering
 \begin{tabular*}{0.5\textwidth}{c c c c c}
 & $\alpha$ [-] &    TI [\%]   &    $\sigma_v$  & $\sigma_w$       \\
 \hline
I1 & 0.08  &    6.1     & 0.71 $\sigma_u$ & 0.64 $\sigma_u$         \\

I2 & 0.13 &    9.3    & 0.65 $\sigma_u$ & 0.57   $\sigma_u$         \\     
    \hline
    \end{tabular*}
\end{table}

\subsection{Computational domain and numerical parameters set-up}

\subsubsection{VFS-Wind}

\begin{figure}
    \centering
    \includegraphics[width=\textwidth]{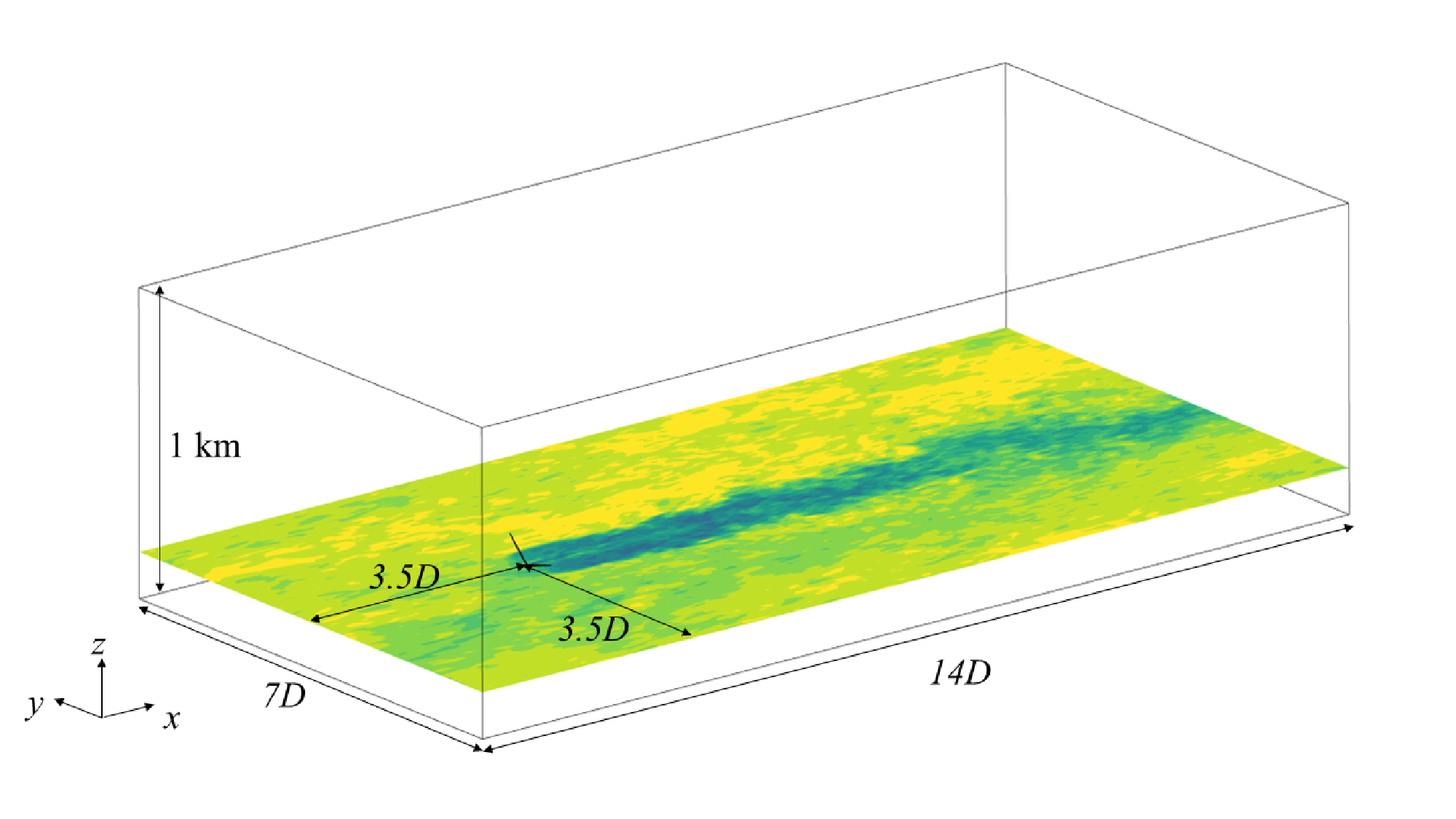}
    \caption{Computational domain of the large eddy simulation in VFS.}
    \label{fig:domain_VFS}
    % F:/work/2023文章/NTNU/computationalDomain.pptx
\end{figure}

The computational domain employed for the large eddy simulation of wind turbine wakes is illustrated by Figure \ref{fig:domain_VFS}. The size of the computational domain is $L_x \times L_y \times L_z = 14D \times 7D \times 1~\textrm{km}$, in streamwise ($x$), transverse ($y$), and vertical ($z$) directions, respectively. The origin of the coordinates coincides with the wind turbine footprint on the ground. Turbulent inflows generated by the precursory simulations are prescribed at the inlet ($x = -3.5D$). At the outlet  ($x = 10.5D$), Neumann boundary condition  is applied to all velocity components ($\partial u_i/\partial x=0$). At the lateral and the top boundaries, the free-slip condition is applied. The bottom boundary employs the same rough-wall function as in the precursor simulations. The domain is discretized by a Cartesian grid with grid nodes of $N_x \times N_y \times N_z = 281\times 281\times 105 $. The grid is uniform in the $x,y$-directions with grid spacing $\Delta x= D/20$ and $\Delta y = D/40$. In the $z$ direction, the grid is uniform near the ground ($z \in (0,2D)$) with $\Delta z  = D/40$ and is gradually stretched to the top boundary. The time step is fixed at $\Delta t = 0.06$ s.  This discretization has been shown to produce grid-independent turbulence statistics of wind turbine wake by a recent study on a similar case \cite{li2021}.

\color{black}

\subsubsection{FAST.Farm}

\begin{figure}[h!]

  \includegraphics[trim={0 0cm 0 0cm},clip,width=0.98\textwidth]{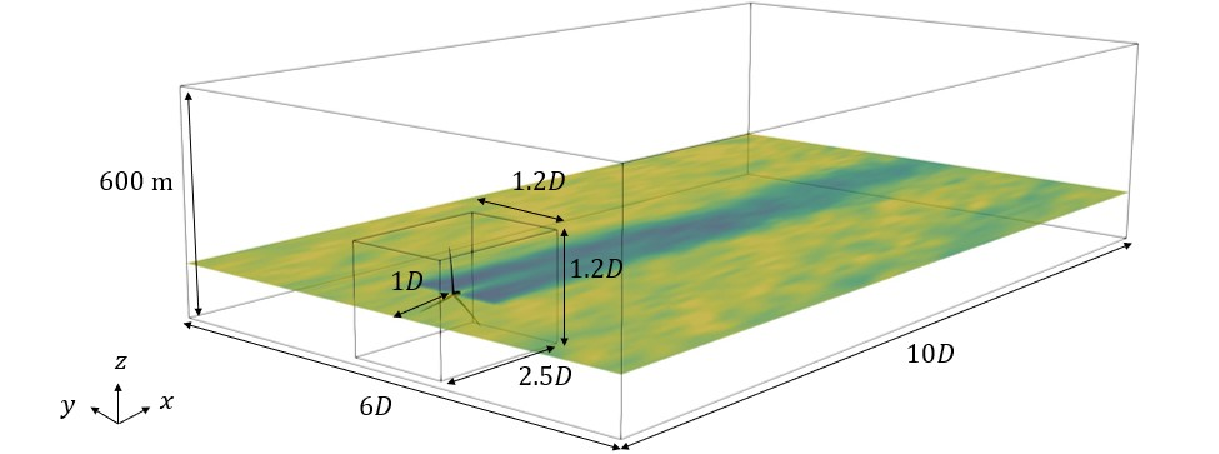}

\centering
\caption{\label{fig:FFdomain}High- and low-resolution computational domains in FAST.Farm. }

\end{figure}

The numerical domain in FAST.Farm, outlined in Figure~\ref{fig:FFdomain}, is $10D$ long, $6D$ wide and $2.5D$ high. Two sub-domains are defined: the larger domain, which corresponds to the low-resolution domain, and the smaller domain around the rotor, which is the high-resolution domain. The low-resolution domain has the same dimensions as the entire domain. The turbine is placed at $x = 1D$, inside the high-resolution domain, which is $1.2D$ wide and high, and it extends over $2.5D$ in the longitudinal direction. The resolution of each domain, based on the work of Shaler et al. \cite{shaler2019}, is presented in Table~\ref{tab:dimensions}, where dSLow is the resolution in $y,z$-directions in the low-resolution domain, and dSHigh the corresponding one in the high-resolution domain. The time steps for both domains, dtLow and dtHigh, are presented in Table~\ref{tab:dimensions}. The recommended values by Shaler et al. and specified in the FAST.Farm user manual \cite{jonkman2021} are indicated as \textit{Rec.}, whereas the ones used in this work are the \textit{Used} ones. The time step in the high-resolution domain is determined by $f_{max}$, which indicates the highest frequencies influencing the structural excitation. $c_{max}$ is the maximum blade chord length of the turbine, which defines the spatial resolution in $y,z$-directions of the high-resolution domain, dSHigh. In the current work, the chosen resolution is above the recommended value of 5.8\,m. However, since the focus of this research is not on the structural analysis of the wind turbine, and also based on the resolution of the LES data, a dSHigh of 8\,m is considered to be sufficient.

\begin{table}[h!]

\caption{Recommended and used values for the spatial and time resolution in the FAST.Farm numerical domain. $f_{max}$ is the highest frequency influencing the structural response and $c_{max}$ the maximum blade chord length. }

\centering
\label{tab:dimensions}

 \resizebox{1\linewidth}{!}{\begin{tabular}{c c c c | c c c c | c c c c }

  \multicolumn{2}{c}{dtLow [s]} & \multicolumn{2}{c|}{dtHigh [s]} & \multicolumn{2}{c}{dxLow [m]} & \multicolumn{2}{c|}{dSLow [m]} & \multicolumn{2}{c}{dxHigh [m]} & \multicolumn{2}{c}{dSHigh [m]} \\

\hline
 
  \textit{Rec}.& \textit{Used} & \textit{Rec.} & \textit{Used} & \textit{Rec.}& \textit{Used} & \textit{Rec.}& \textit{Used} & \textit{Rec.}& \textit{Used} & \textit{Rec.}& \textit{Used} \\
 \hline
 $< \dfrac{C_{meand}  D_{w}}{10U_{\infty,z_{HH}}}$ &    & $ <  \dfrac{1}{2f_{max}}$  &    &     &  &  $ < \dfrac{C_{meand} D_{w}U_{\infty,z_{HH}}}{150\text{m/s}}$   &    &  &   &  $<c_{max}$ &   \\
< 5.0 &  3.0  & < 0.5  & 0.2   &  0.11$D$  & 0.11$D$   &  < 0.12$D$  &  0.10$D$   &    &  0.0075$D$ & < 5.8  & 8.0   \\

\hline

\end{tabular}}

\end{table}

%\subsection{Wind turbine parametrization method}

\section{Results}\label{sec:results}

\subsection{Time averaged wake}

\subsubsection{Horizontal wake steering}

\begin{figure}[h!]
\begin{subfigure}[b]{1\textwidth}
\includegraphics[trim={0 0cm 0 0cm},clip,width=1\textwidth]{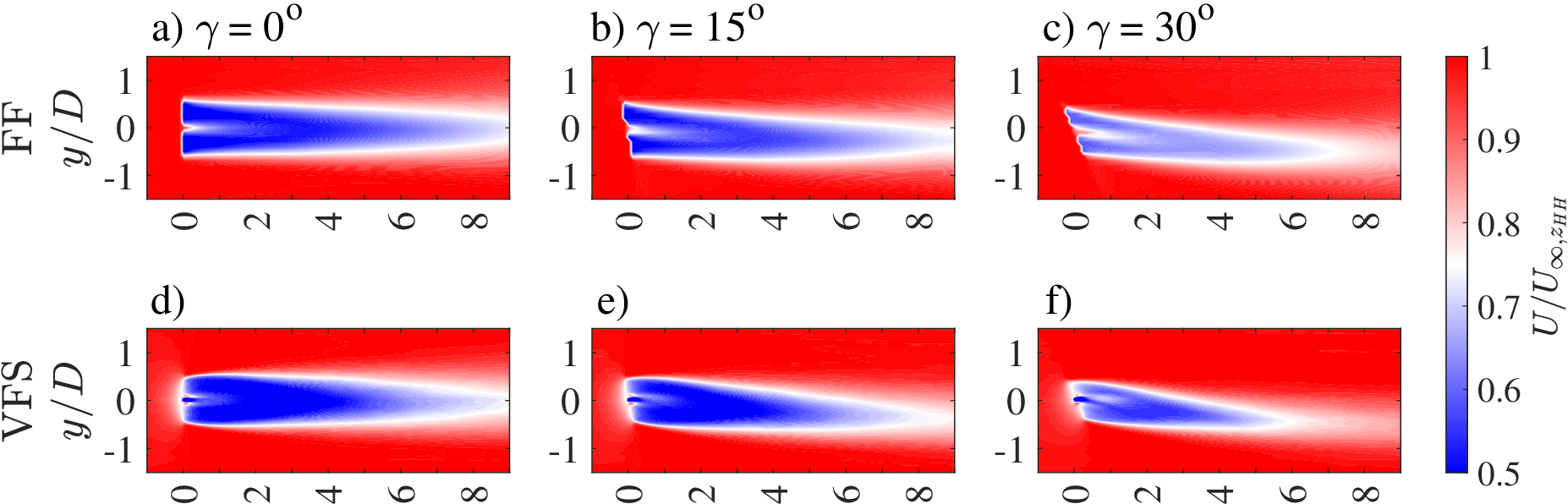}
\end{subfigure}
\hfill

\begin{subfigure}[b]{1\textwidth}
\includegraphics[trim={0 0cm 0 0cm},clip,width=1\textwidth]{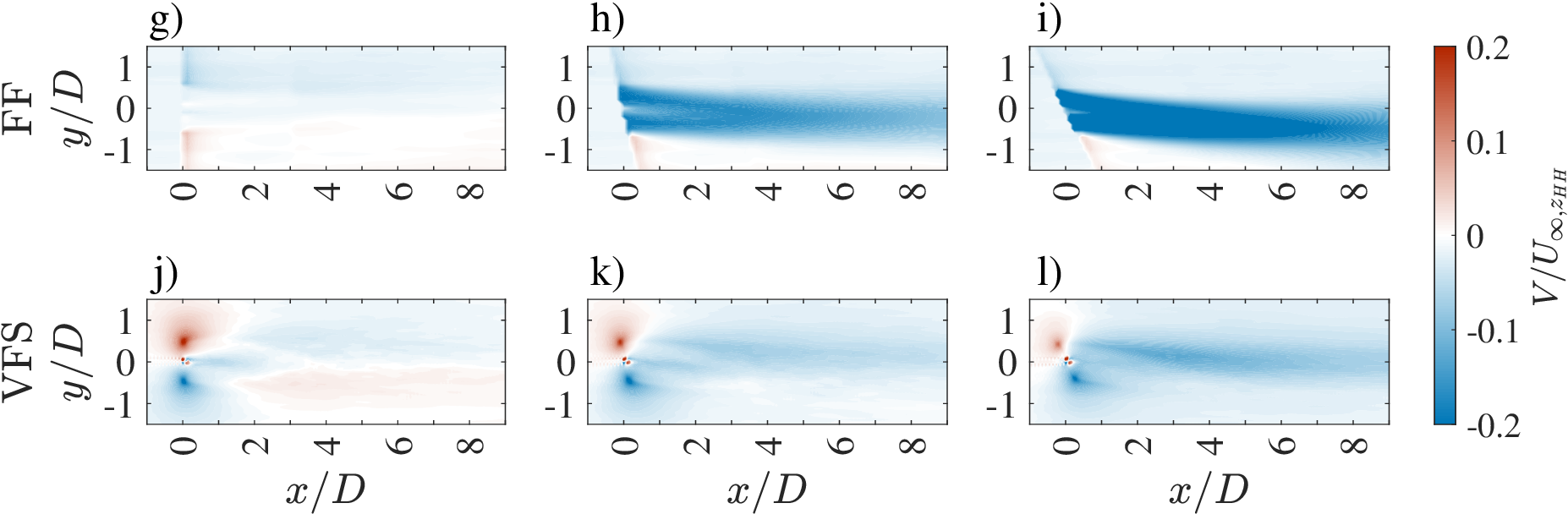}
\end{subfigure}
\caption{ \label{fig:wakeDeficitXYfig4} Time-averaged velocity on the horizontal ($xOy$) plane at hub-height for three representative yaw deflection angles $\gamma = 0^\circ,~ 15^\circ, ~30^\circ$, for inflow 1; a-c) streamwise velocity deficit $U/U_{\infty,z_{HH}}$ computed by FAST.Farm, and d-f) by VFS; g-i) horizontal velocity $V/U_{\infty,z_{HH}}$ computed by FAST.Farm, and j-l) by VFS. }
\end{figure}

The predictive capability of FAST.Farm for cases with yaw misalignment is first examined by the time-averaged contour plots at the hub-height plane for three representative yaw angles, $\gamma= 0$\textdegree, 15\textdegree\xspace and 30\textdegree, as shown in Figure \ref{fig:wakeDeficitXYfig4}. 

Figure~\ref{fig:wakeDeficitXYfig4} a) to c) presents the streamwise component of the wind velocity $U$ predicted by FAST.Farm (FF). Figure~\ref{fig:wakeDeficitXYfig4} d) to f) shows the same results, as obtained with VFS-Wind. The streamwise velocity predicted by both approaches is similar. For all yaw angles, both models provide consistent streamwise development of the wake, characterised by the expansion of the wake and the recovery of streamwise velocity due to the turbulent mixing and wake meandering mechanism \cite{porte2020wind}. As the yaw misalignment angle increases, the wake deficit and the width of the low-speed region both decrease for both models, due to the decrease of thrust, and of the projected rotor area. In both models, the streamwise component of the velocity field is approximately symmetric with respect to the wake centerline. 

Some minor discrepancies between the models are also observed. The first difference between the two models for a zero yaw angle is at the near-wake region, as shown in Figure \ref{fig:wakeDeficitXYfig4} a) and d). In FF, the wake width starts to decrease linearly right behind the rotor location ($x=$0). In contrast, the near wake region predicted by VFS has a constant width up to almost two rotor diameters downstream ($x=2D$). From that point on, the decrease of the wake deficit and the width of the low-speed region is slightly slower for the FF model. The same trend is also observed in $\gamma= 15$\textdegree\xspace and 30\textdegree\xspace, when comparing  Figure~\ref{fig:wakeDeficitXYfig4} b-c) and e-f).  

However, the lateral velocity component ($V$) in the wake predicted by both approaches shows a larger discrepancy, as shown in Figure~\ref{fig:wakeDeficitXYfig4} g-i) and j-l).  
%This symmetry is not observed for the lateral component $V$, as shown in Figure~\ref{fig:wakeDeficitXYfig4} g-i) and j-l) for FF and for the VFS model, respectively. 
For the case with $\gamma = 0 ^\circ$, the time-averaged transverse velocity $V$ is symmetric to the wake centreline and is almost zero for both cases. In the far-wake, both models predict a converging pattern of the transverse velocity, reflecting the mean convection from the freestream to the wake that contributes to the wake recovery. It is worth noting that VFS predicts a diverging transverse velocity at the rotor position $x=0$ in Figure \ref{fig:wakeDeficitXYfig4} j), which reflects the initial wake expansion enforced by the mass conservation \cite{burton2011wind}. Such an initial wake expansion is not captured by FF, since its governing equation for the wake deficit (Eq. \eqref{eq:deficitEq}) is simplified with the far-wake assumption without imposing the continuity condition. As the yaw misalignment angle increases, so does the transverse component, and the symmetry with respect to the centreline vanishes. Furthermore, the induced velocity increases in the negative $y$-direction, and its magnitude increases with $\gamma$.  Figure~\ref{fig:wakeDeficitXYfig4} h), and especially i), indicates an overestimation of the mean value of the transverse component in FF with respect to that of VFS. Besides the magnitude, the transverse components predicted by both models reside in different locations. In the result of FF, $V$ reside in the same region with the wind speed deficit. However, the present LES results as well as previous experiments \cite{bastankhah2016} and simulations \cite{li2021} suggest that the transverse velocity should be skewed towards the wake border behind the rotor's leading edge. This difference indicates that the wake deflection in FAST.Farm is driven only by the transverse velocity component created directly in the near wake, with other physics leading to the wake centerline deflection, e.g., the wake deformation \cite{howland2016wake}, being ignored.

%The induced transverse component in the VFS model could be explained not only by the transverse induced velocity due to the non-zero transverse thrust, but by the expansion of the wake itself. 

\begin{figure}[h!]
\centering
   \sbox0{\solidCyan}\sbox1{\solidRed} \sbox2{\solidGreen}\sbox3{\solidPurple}\sbox4{\solidYellow}\sbox5{\solidLine}\sbox6{\dashLine}%
    \centering
    \framebox{\shortstack{\usebox0   $\gamma$=0\textdegree \xspace, \usebox1 $\gamma$=10\textdegree \xspace,
    \usebox2   $\gamma$=15\textdegree \xspace, \usebox3   $\gamma$=20\textdegree \xspace,\usebox4 $\gamma$=30\textdegree \xspace\\
    \usebox5   FAST.Farm \xspace, \usebox6 VFS-Wind} }
\centering
\vspace{5mm}

\begin{subfigure}[b]{1\textwidth}
\includegraphics[trim={0 14cm 0 0},clip,width=1\textwidth]{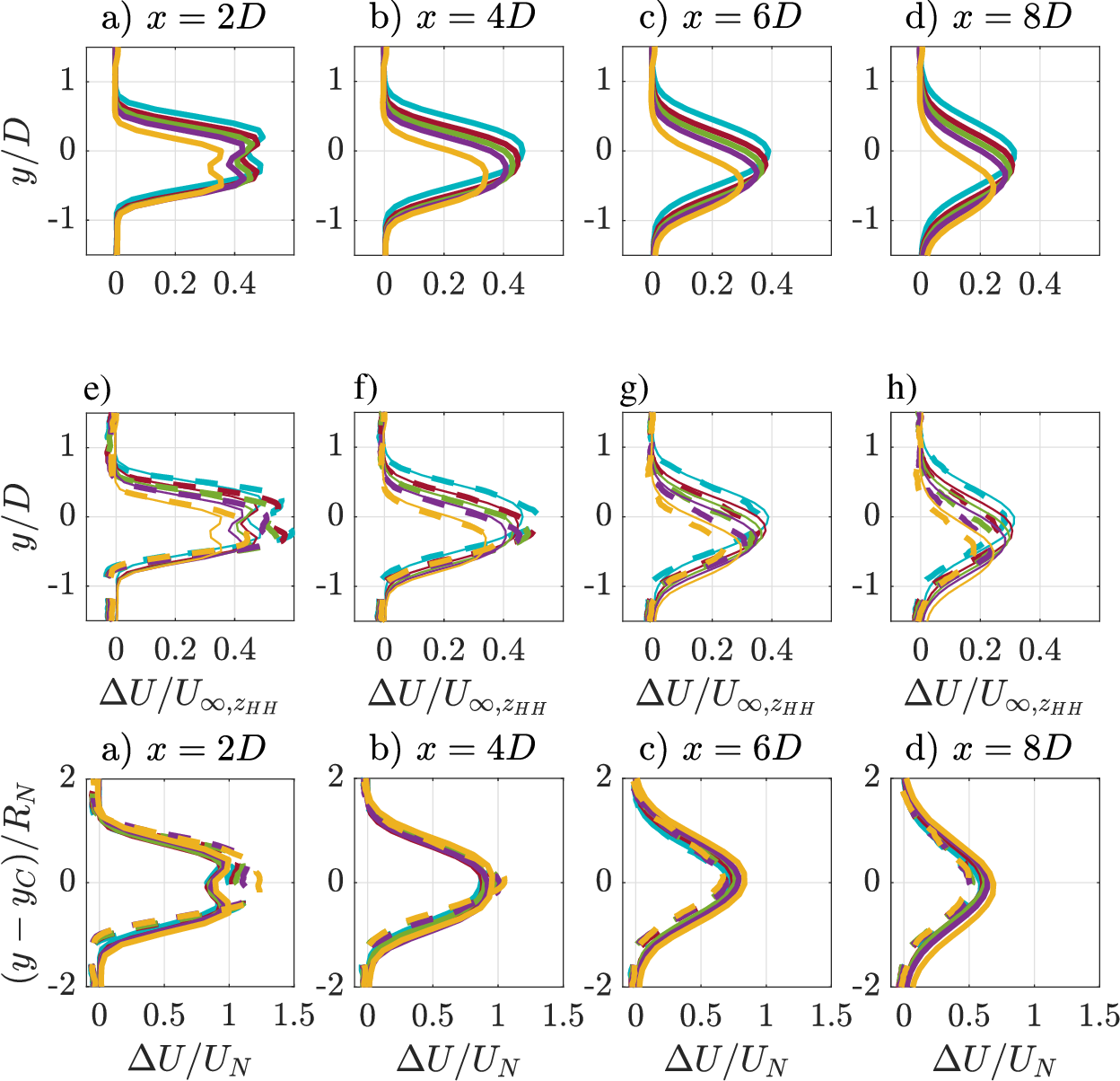}
\end{subfigure}
\hfill
\begin{subfigure}[b]{1\textwidth}
\includegraphics[trim={0 6.6cm 0 6.8cm},clip,width=1\textwidth]{./Figures/MS_deficitPolarXYwithLines.eps}
\end{subfigure}
%\hfill
%\begin{subfigure}[b]{1\textwidth}
%\includegraphics[trim={0 0 0 8.25cm},clip,width=1\textwidth]{./Figures/MS_deficitPolarXY.eps}
%\end{subfigure}
\caption{ \label{fig:wakeDeficitXY} Time-averaged velocity deficit $\Delta U(y)/U_{\infty,z_{HH}}$ at $x = 2,~ 4,~ 6$ and $8D$, at the hub-height plane for the five yaw deflection angles $\gamma$, for inflow 1; a-d) is computed by FAST.Farm, and e-h) by VFS, with dashed lines. The thin lines show the FAST.Farm results (same as a) to d), for the simplicity of the comparison.}
\end{figure}

In Figure \ref{fig:wakeDeficitXY}, the time-averaged wake is examined in further detail. Figure~\ref{fig:wakeDeficitXY} a) to h) presents the wake deficit $\Delta U(y)/U_{\infty,z_{HH}}$, with $\Delta U (y) = U(y)-U_{\infty,z_{HH}}$, at the hub-height. The results are presented at four downwind locations, $x= 2$,~ 4,~ 6 and 8$D$, for the yawed rotor with five $\gamma$ angles. The $C_\text{meand}$ filter used in the results shown in this section are the ones presented in Table~\ref{tab:cMeand}, cases 0 to 4a. The solid lines, in the top row (a to d), represent the deficit as computed in FAST.Farm. The deficits in the VFS model, with dashed lines, are shown in the bottom row (e to h), together with the FAST.Farm results, i.e. the same as shown in the top row (a to d), outlined by thinner lines, to make the comparison between the models more clear. Similar trends are observed for both models: both the wake width and deficit decrease as the yaw angle increases. If the deficit between VFS and FAST.Farm is compared, the latter predicts a higher deficit at the near-wake region, i.e. at $x= 2D$. As the downstream distance $x$ increases, the wake deficits are similar. 

Li and Yang \cite{li2021} suggested that there is a similarity of wakes for turbines with different yaw angles based on large eddy simulations of a 2.5 MW wind turbine. Based on that work, the time-averaged wake velocity deficit profiles $\Delta U(y)$ for cases with different yaw angles collapse if $\Delta U(y)$ is normalised by a proper velocity ($U_N$) and length ($R_N$) scale, derived based on one-dimensional momentum theory in the streamwise direction. The characteristic velocity and length are defined as follows:

\begin{equation}
U_N = U_{\infty,z_{HH}}\left(1-\sqrt{(1-\tilde{C}_T\cos ^2 \gamma)}\right ),
\end{equation}
and
\begin{equation}
R_N=R \cos \gamma \sqrt{\left(1+\sqrt{1-\tilde{C}_T \cos ^2 \gamma}\right) \bigg/\left(2 \sqrt{1-\tilde{C}_T \cos ^2 \gamma}\right)}.
\end{equation}

\begin{figure}[h!]
\centering
   \sbox0{\solidCyan}\sbox1{\solidRed} \sbox2{\solidGreen}\sbox3{\solidPurple}\sbox4{\solidYellow}\sbox5{\solidLine}\sbox6{\dashLine}%
    \centering
    \framebox{\shortstack{\usebox0   $\gamma$=0\textdegree \xspace, \usebox1 $\gamma$=10\textdegree \xspace,
    \usebox2   $\gamma$=15\textdegree \xspace, \usebox3   $\gamma$=20\textdegree \xspace,\usebox4 $\gamma$=30\textdegree \xspace\\
    \usebox5   FAST.Farm \xspace, \usebox6 VFS-Wind} }
\centering
\vspace{5mm}

\includegraphics[trim={0 0 0 13.5cm},clip,width=1\textwidth]{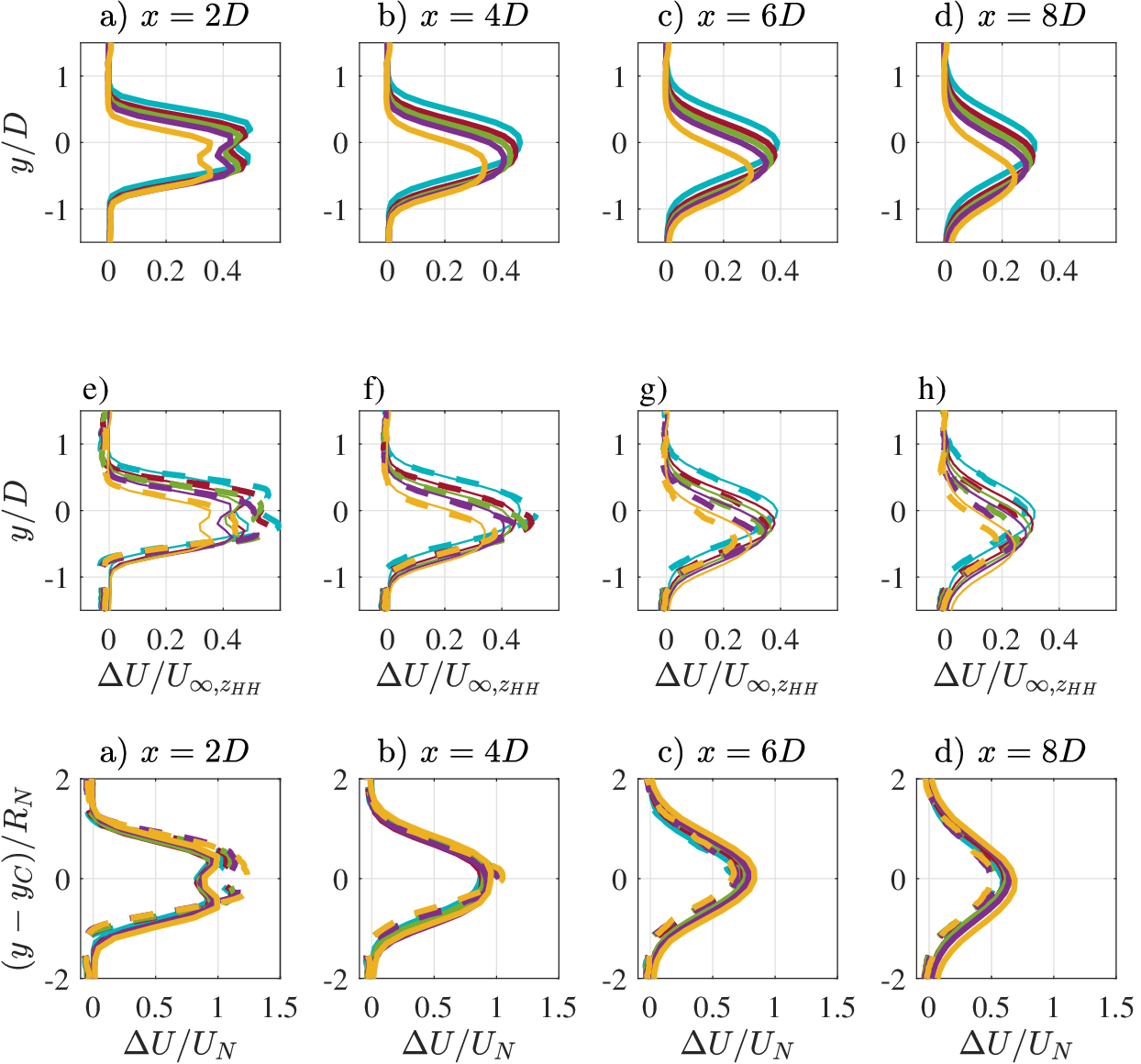}

\caption{ \label{fig:wakeDeficitXYNorm} Time-averaged velocity deficit at $x = 2,~ 4,~ 6$ and $8D$, at the hub-height plane for the five yaw deflection angles $\gamma$, for inflow 1. The profiles are normalized by $U_N$, and the wake centers shifted with respect to $y_c$ and normalized by the length scale $R_N$.}
\end{figure}

With the above characteristic scales, the results for different yaw angles predicted by both methods can be compared quantitatively, as in Figure~\ref{fig:wakeDeficitXYNorm}. Figure~\ref{fig:wakeDeficitXYNorm} a) to d) shows the deficit shifted by the wake centers $y_c$, and normalised by the characteristic velocity $U_N$ and $R_N$. A good agreement between the models is observed in the normalised transverse profiles of the streamwise component beyond the near wake region $x = 2D$. This collapse also indicates that the velocity deficit profiles are almost symmetric with respect to the wake centreline for every yaw misalignment angle and the similarity of wake for wind turbines operating at different yaw angles is well captured by both FAST.Farm and VFS-Wind. A small discrepancy between the normalized wake deficit profiles is observed, where the wake deficit predicted by FAST.Farm is larger than VFS, in the far-wake ($x = 8D$), consistent with the observation from the contour plots in Figure \ref{fig:wakeDeficitXYfig4}.   
 Although these results are specific to inflow 1, i.e. with a lower TI level, the same comparative trends between the VFS and FAST.Farm results are also observed for inflow 2, and are not shown for the sake of conciseness. 

\subsubsection{ Vertical wake steering}

\begin{figure}[h!]
\begin{subfigure}[b]{1\textwidth}
\includegraphics[trim={0 0cm 0 0cm},clip,width=1\textwidth]{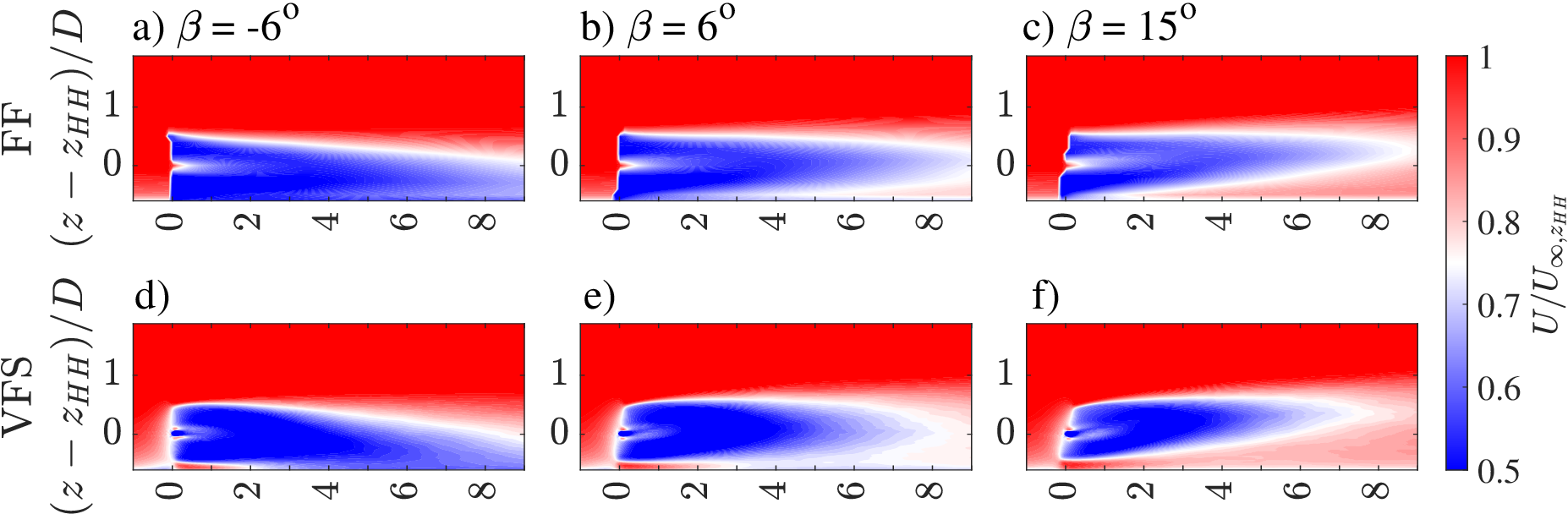}
\end{subfigure}
\hfill

\begin{subfigure}[b]{1\textwidth}
\includegraphics[trim={0 0cm 0 0cm},clip,width=1\textwidth]{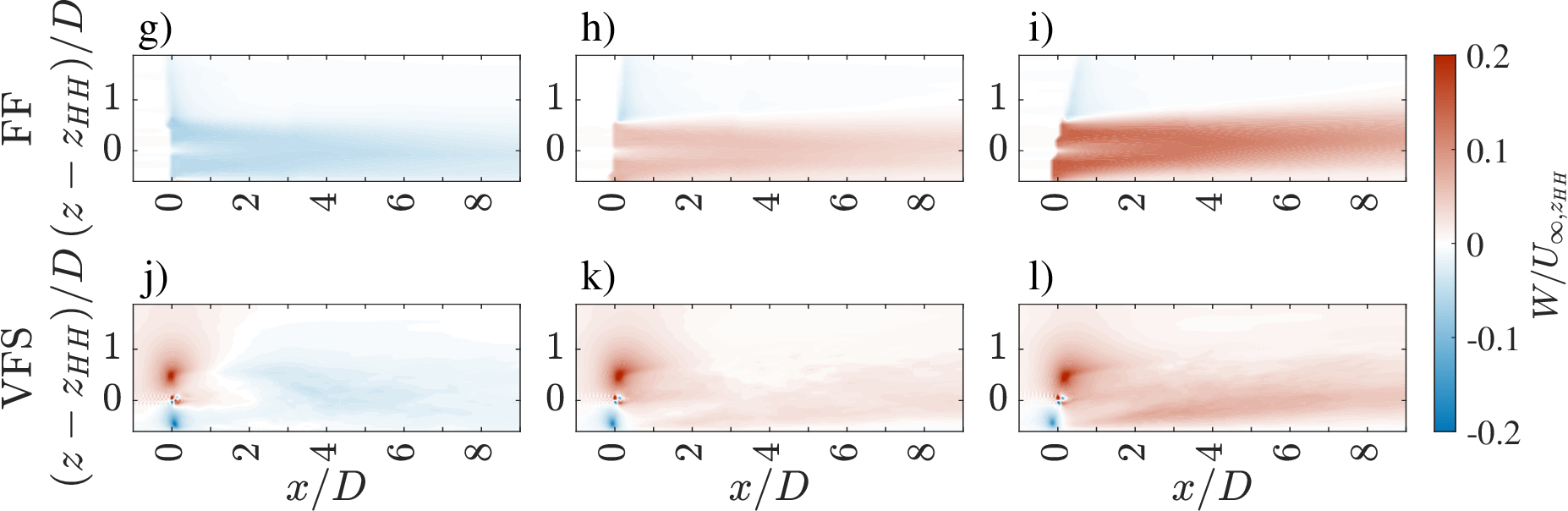}
\end{subfigure}
\caption{ \label{fig:wakeDeficitXZfig4} Time-averaged velocity on the vertical ($xOz$) plane for three representative tilt deflection angles $\beta = -6^\circ,~ 6^\circ, ~15^\circ,$ for inflow 1; a-c) streamwise velocity deficit $U(z)/U_{\infty,z_{HH}}$ computed by FAST.Farm, and d-f) by VFS; g-i) vertical velocity  $W(z)/U_{\infty,z_{HH}}$ computed by FAST.Farm, and j-l) by VFS.  }
\end{figure}

Figure~\ref{fig:wakeDeficitXZfig4} a) to c) presents the time-averaged longitudinal component of the wind speed $U(z)$, normalized by the undisturbed incoming wind speed at $z_{HH}$, denoted as $U_{\infty,z_{HH}}$, at the vertical ($xOz$) plane for three tilt misalignment angles, namely $\beta = -6^\circ,$ $6^\circ$ and 15\textdegree, for FAST.Farm. Figure~\ref{fig:wakeDeficitXZfig4} d) to f) shows the analogous results, but for VFS. In general, the vertical deflection predicted by FAST.Farm and VFS  shows a good agreement. For the negative tilt angle ($\beta = -6^\circ$), the wake deficit is deflected downwards as the distance downstream increases, whereas the wake is deflected upwards for cases with positive tilt angles, and the wake width and velocity deficit both decrease with increasing tilt angle similar to the cases with yaw misalignment. Similar to the observations for the yaw misalignment cases (Figure \ref{fig:wakeDeficitXYfig4}), the initial wake width does not change until approximately $x=2D$ for the VFS model, whereas a linear decrease of the width of the low wind speed region is observed in FAST.Farm throughout the entire wake region. In the far-wake, the wake deficit in the VFS model is recovered at a higher rate, reflecting a stronger exchange of momentum with freestream flow predicted by the high-fidelity model. The vertical velocity component $W(z)/U_{\infty,z_{HH}}$ is shown in Figure~\ref{fig:wakeDeficitXZfig4} g-i) and j-l) for FF and for VFS, respectively. For the negative tilt angle case, the vertical component is a small negative value, for both models. For the positive tilt angles, this component is positive. Again, the vertical velocity predicted by FF model is stronger than that predicted by VFS. For the FF prediction, this vertical velocity originates from the rotor location, and can be explained by the  non-zero thrust component in the vertical direction. In comparison, the vertical component in the VFS model has a smaller magnitude, from the beginning of the wake indicating that the vertical component of the rotor thrust is not the only driving mechanism leading to a vertical wake deflection in large eddy simulation.

\begin{figure}[h!]
\centering
    \sbox0{\solidLine}\sbox1{\solidCyan}\sbox2{\solidRed} \sbox3{\solidGreen}\sbox4{\solidPurple}\sbox5{\solidYellow}\sbox6{\solidLine}\sbox7{\dashLine}%
    \centering
    \framebox{\shortstack{\usebox0   $\beta$=-6\textdegree \xspace, \usebox1 $\beta$=0\textdegree \xspace,
    \usebox2   $\beta$=6\textdegree \xspace, \usebox3   $\beta$=10\textdegree \xspace,\usebox4 $\beta$=15\textdegree \xspace,\usebox5 $\beta$=20\textdegree \xspace\\
    \usebox6   FAST.Farm \xspace, \usebox7 VFS-Wind} }
\centering
\vspace{5mm}

\begin{subfigure}[b]{1\textwidth}
\includegraphics[trim={0 13.65cm 0 0},clip,width=1\textwidth]{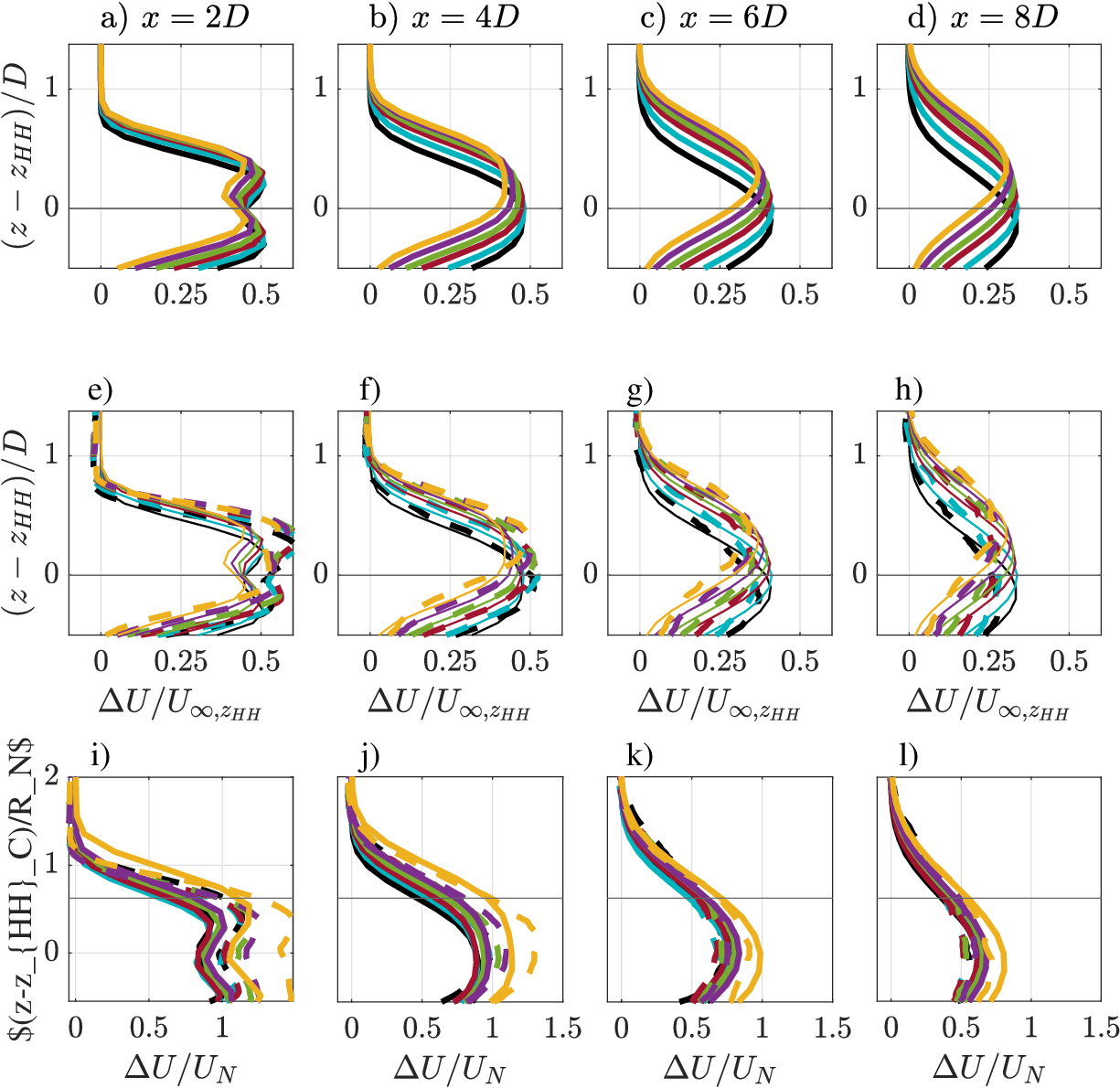}
\end{subfigure}
\hfill
\begin{subfigure}[b]{1\textwidth}
\includegraphics[trim={0 6.5cm 0 6.5cm},clip,width=1\textwidth]{./Figures/MS_deficitPolarXZLines.eps}
\end{subfigure}
\caption{ \label{fig:wakeDeficitXZ} Time-averaged velocity deficit $\Delta U (z)/U_{\infty,z_{HH}}$ at $x = 2,~ 4,~ 6$ and $8D$, for $-0.5D<(z-z_{HH})<1.3D$, at the rotor plane for the six tilt angles $\beta$, for inflow 1; a-d) is computed by FAST.Farm, and e-h) by VFS. The hub-height is at $(z-z_{HH})/D = 0$, outlined by the solid horizontal line. The thin lines show the FAST.Farm results (same as a) to d), for the simplicity of the comparison. %{\color{red} Irene: how $\Delta U$ is defined, does it defined as $\Delta U(z) = U(z)-U_\infty$ or $\Delta U(z) = U(z)-U_\infty$.} 
}
\end{figure}

Figure~\ref{fig:wakeDeficitXZ} a) to h) presents the wake deficit $\Delta U (z)/U_{\infty,z_{HH}}$, with $\Delta U (z) = U(z)-U_{\infty,z_{HH}}$, at $x = 2,~ 4,~ 6$ and $8D$, for the six tilt deflection angles. The $C_\text{meand}$ filter used in the results shown in this section are the ones presented in Table~\ref{tab:cMeand}, cases 1b to 5 (and 0 for the zero tilt angle case). The solid lines, in the top row, a to d), show the deficit from FAST.Farm. The deficits in the VFS model, with dashed lines, are shown in the bottom row (e to h), together with the FAST.Farm results, i.e. the same as shown in the top row (a to d), outlined by thinner lines, to make the comparison between the models more clear. The vertical profiles have a non-symmetric shape, due to the ground limit at $z = 0$. The maximum deficit point deflects from the hub-height center position, i.e. $(z-z_{HH})/D = 0$, as it moves further downstream. If the tilt angle $\beta$ is positive, the maximum deficit position deflects upwards, whereas for $\beta = -6$\textdegree\xspace, the maximum deficit deflects downwards. The shape follows a Gaussian function, except for the $\beta = -6^\circ$ and 0\textdegree\xspace cases. At $x = 8D$, the FAST.Farm model generally overpredicts the deficit, reaching a maximum difference of 24\% between the models.

\subsection{ Wake meandering}\label{sec:meand}

\begin{figure}[h!]
\begin{subfigure}[b]{1\textwidth}
\includegraphics[trim={0 0cm 0 0cm},clip,width=1\textwidth]{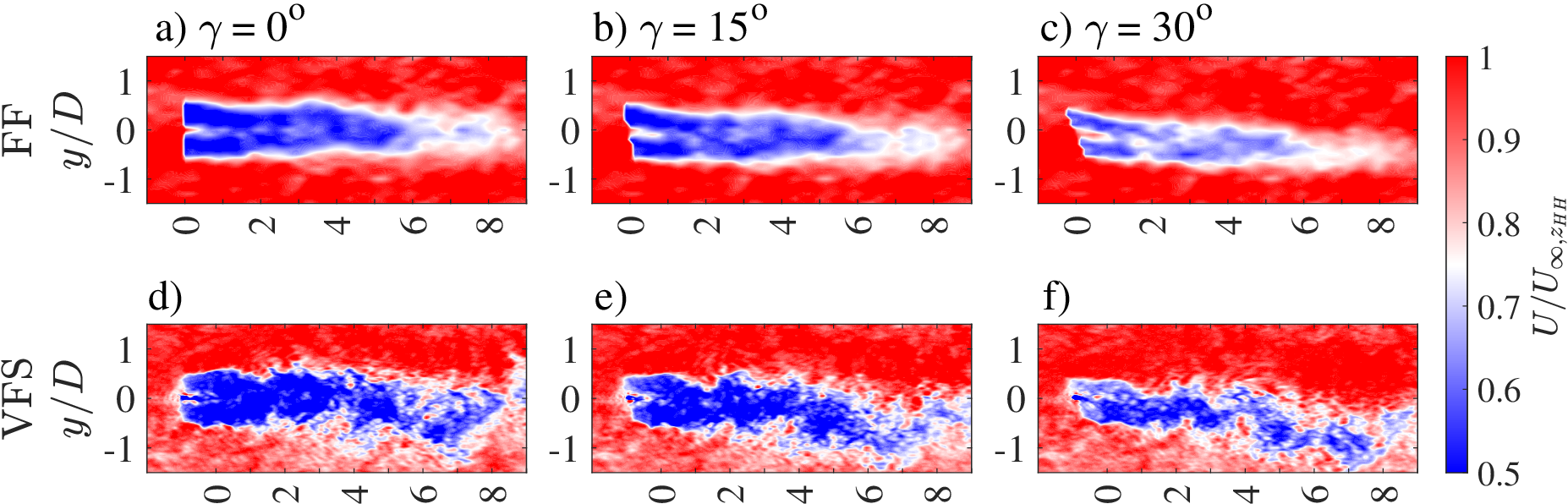}
\end{subfigure}
\hfill

\begin{subfigure}[b]{1\textwidth}
\includegraphics[trim={0 0cm 0 0cm},clip,width=1\textwidth]{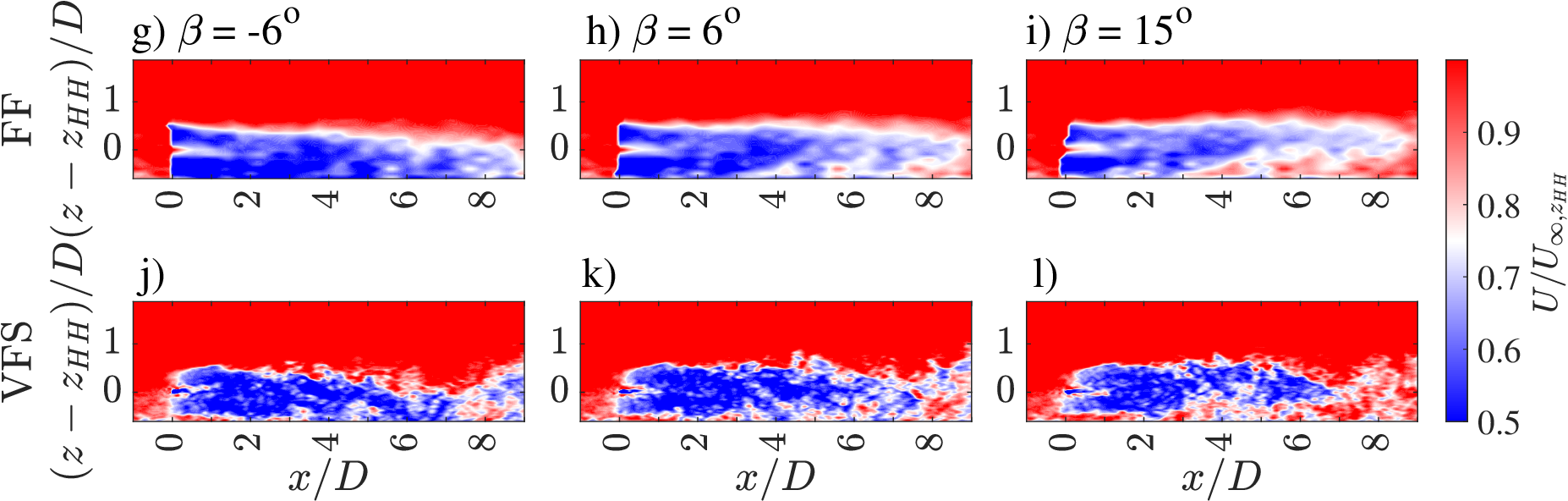}
\end{subfigure}
\caption{ \label{fig:wakeDeficitXYfig4Inst} Instantaneous flow field for $\gamma = 0$\textdegree, 15\textdegree\xspace and 30\textdegree\xspace and $\beta = -6$\textdegree, 6\textdegree\xspace and 15\textdegree, for inflow 1; a-c) streamwise velocity deficit $U(y)/U_{\infty,z_{HH}}$ at the hub-height plane $xOy$ computed by FAST.Farm, and and d-f) by VFS; g-i) streamwise velocity deficit $U(z)/U_{\infty,z_{HH}}$ at the $xOz$ plane computed by FAST.Farm, and d-f) by VFS. }
\end{figure}

In this subsection, we investigate the predictive capability of wake meandering of FAST.Farm against the results of large eddy simulation. Figure~\ref{fig:wakeDeficitXYfig4Inst} presents the instantaneous flow field $U(y)/U_{\infty,z_{HH}}$ at the hub-height plane, at the same instant, for $\gamma=$ 0\textdegree, 15\textdegree\xspace and 30\textdegree\xspace and for $\beta=-$6\textdegree, 6\textdegree\xspace and 15\textdegree, for both FAST.Farm and VFS, for inflow 1. As seen, the instantaneous wake contours predicted by both approaches show similar silhouettes. However, the instantaneous wake predicted by VFS contains more turbulent structures at smaller scales, which are not resolved by the DWM model. Based on these instantaneous wake fields, we define the instantaneous wake center position ($y_c(x)$ or $z_c(x)$) at a downstream location $x$ as: the position where the maximum instantaneous wake deficit is located, for FAST.Farm; and the weight center of the velocity deficit, for VFS.

\subsubsection{Statistics of the instantaneous wake center}

\begin{figure}[h!]
\centering
   \sbox0{\scircle{3pt}}\sbox1{\striangle} \sbox2{\solidPurple}\sbox3{\solidYellow}%
    \centering
    \framebox{\shortstack{\usebox0   FAST.Farm, \usebox1 VFS-Wind\\
    \usebox2   inflow 1  , \usebox3 inflow 2} }
\centering

\vspace{0.5cm}

\includegraphics[width=0.98\textwidth]{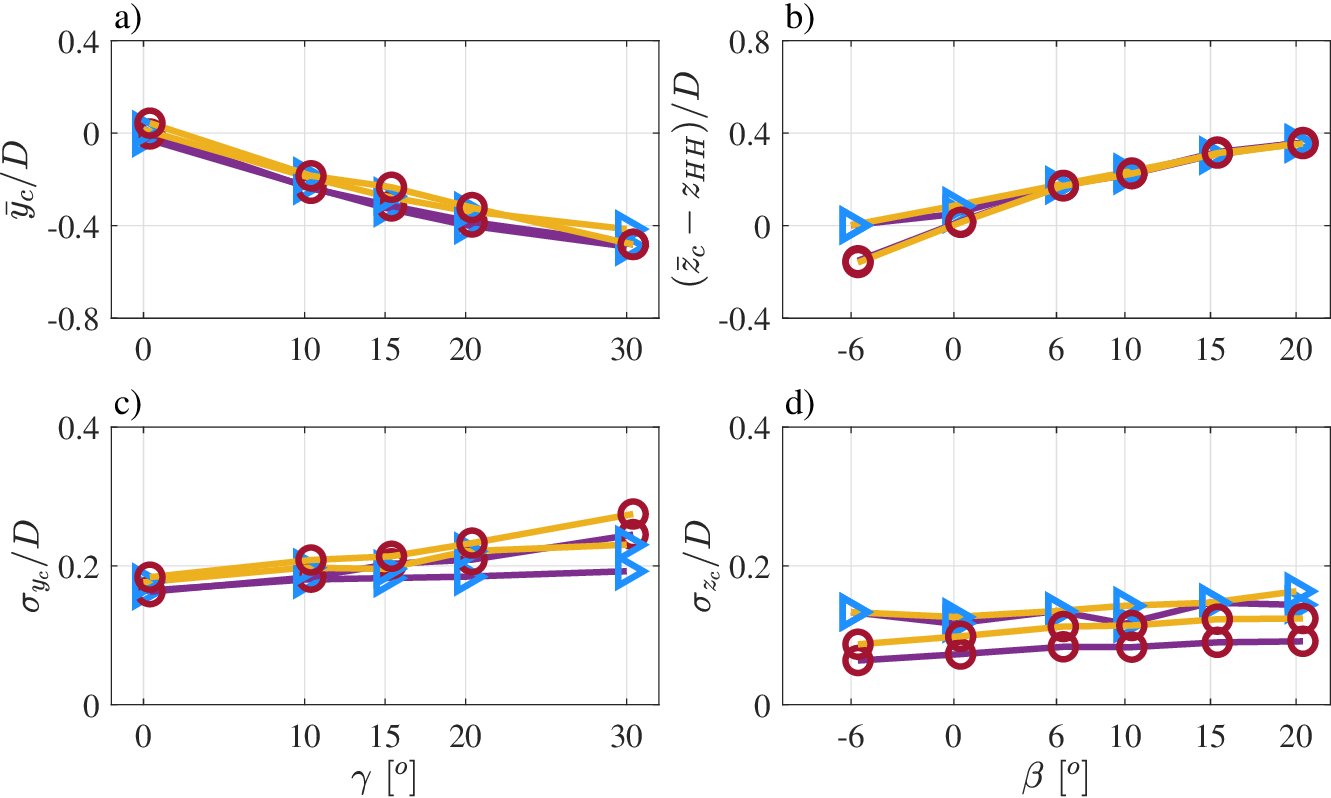}
\caption{ \label{fig:IEAComparisonMeand} Mean horizontal and vertical position, $\bar{y}_c$ and $\bar{z}_c$ (a, b) and standard deviation $\sigma$ (c, d) of the wake center position at $8D$ for the yaw and tilt misalignment cases, respectively, for inflow 1 and inflow 2. The hub-height $z_{HH}$ is subtracted from the mean vertical deflection.}
\end{figure}

Figure~\ref{fig:IEAComparisonMeand} presents an overview of the mean and standard deviation of the horizontal and vertical positions of the wake center, $y_c$ and $z_c$, respectively, of the wake, at $x=8D$. The mean and standard deviation of the horizontal deflection is shown for the yaw misalignment cases, whereas the mean and standard deviation of the vertical deflection is presented for the tilt angle cases. %The wake center here is identified as the instantaneous point of the maximum deficit in FAST.Farm, {\color{red} and as the weight center of the instantaneous velocity deficit in VFS.}  
The statistical values of the wake meandering as computed with FAST.Farm are outlined by red circles, and the corresponding mean values for the VFS model are depicted by blue triangles. Both the mean and standard deviation values are normalized by the rotor diameter. The hub-height $z_{HH}$ is subtracted from the mean vertical deflection $z_c$.

One key finding in comparing the wake-meandering statistics of FAST.Farm and VFS is that the results of FAST.Farm are sensitive to the $C_\text{meand}$ filter. $C_\text{meand}$ has to be adjusted to fit the mean and standard deviation values for each yaw misalignment and tilt deflection case, respectively, to the VFS data. Table~\ref{tab:cMeand} presents the values for $C_\text{meand}$ that yielded the closest mean and standard deviation compared to VFS at $8D$. Interestingly, the optimal $C_\text{meand}$ is found to be different for cases with different yaw and tilt angles. For correctly predicting the horizontal wake meandering with wind turbine yaw, the optimal is found in the range $1.90<C_\text{meand}<2.10$, which is close to the default value $C_\text{meand}=2$ proposed by Larsen et al. \cite{larsen2008}. However, the optimal value for $C_\text{meand}$ is significantly larger for predicting the vertical oscillation of the wake for a tilted rotor.

The largest difference between the models in the mean value of the wake deflection at $x=8D$, for the yaw misalignment case (Figure~\ref{fig:IEAComparisonMeand} a), is for $\gamma = 15$\textdegree\xspace case, for which FAST.Farm overestimates, for I2, the mean value by 15\%. The standard deviation in these cases is overestimated by FAST.Farm, especially for the 4a case ($\beta = 30$\textdegree), where the value is 25\% higher. This overestimation is explained by the larger $C_\text{meand}$. The larger this filter, the larger the polar grid area used to calculate the spatial-averaged velocity with which the wake planes meander. A larger polar grid implies that the wake deficit will affect the longitudinal component less, and the average over the polar grid will be larger. This larger averaged $U$-component will lead to a smaller $V/U$ ratio, and therefore to smaller mean and standard deviation of the wake center. $C_\text{meand} = 2.10$ for the 4a case yields the closest standard deviation $\sigma_{y_c}$ to VFS without compromising the mean horizontal meandering of the wake. The effect of the choice of this filter in FAST.Farm is discussed further in this section and in the following one, section~\ref{sec:Cmeand}. %The turbulence intensity does not affect the mean deflection for either of the models. However, the standard deviation is affected by the turbulence level; the cases with inflow 2 show a larger standard deviation, which is consistent with the higher turbulence intensity and $\sigma_v/U_{\infty,z_{HH}}$, as seen in Figure~\ref{fig:IEAfigure2}. 

%TC:ignore
%{
%\color{red} Hi, Irene, the above paragraph is not clear to me.  Could you add more explanation about the effect of $C_\text{meand}$ in this paragraph? I think it is an important point for the paper. \color{green} I agree it was very unclear. I re-wrote it.
%}
%TC:endignore

The mean value of the vertical meandering is barely affected by the turbulence intensity for either of the models. The standard deviation of the vertical meandering of the wake center $\sigma$\textsubscript{$z_c$}, for the VFS model, is not seen to be impacted by the TI, but it is for the FAST.Farm model. With increasing turbulence intensity, it is found that FF predicts a stronger wake meandering. This difference in $\sigma$\textsubscript{$z_c$} may be related to the ground effect, which is only modelled in large eddy simulations. Furthermore, Figure \ref{fig:IEAComparisonMeand} c-d) shows that for both models, the standard deviation of the vertical meandering is smaller than that of the horizontal meandering, due to the lower standard deviation of the vertical component, $\sigma_w$ with respect to that of the lateral one, $\sigma_v$, which is consistent with the velocity fluctuations of inflow as shown in Figure~\ref{fig:IEAfigure2} c - d). 

%, and the reason for the VFS not being affected by the difference of TI of the inflow may be the increased amount of momentum experienced as the wake advects downstream, which evens out the initial turbulence intensity for the two incoming inflows. This effect is not modelled by the DWM model, and therefore it is expected that the initial turbulence intensity level will affect the meandering downstream. 

\begin{figure}[h!]
\centering
   \sbox0{\scirclee{3pt}}\sbox1{\striangle}
    \framebox{\shortstack{\usebox0  FAST.Farm; \usebox1 VFS-Wind} }
\centering

\vspace{0.5cm}

\includegraphics[width=1\textwidth]{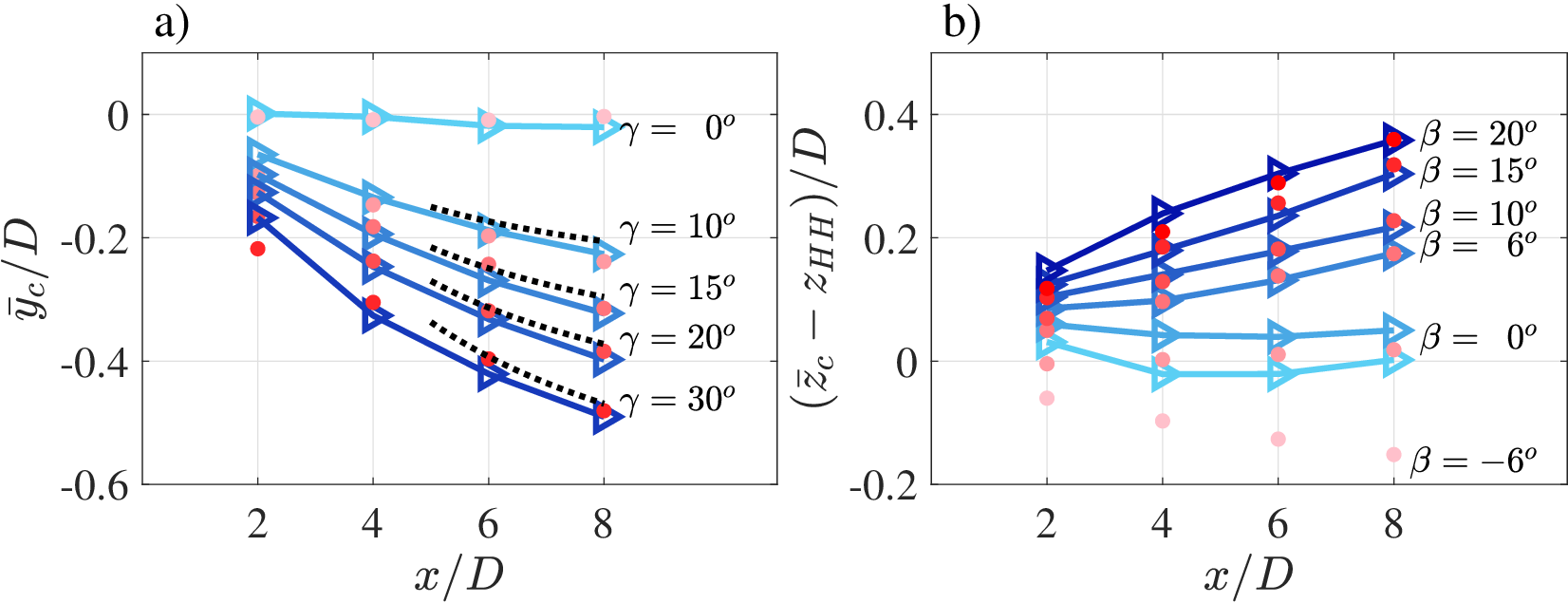}
\caption{ \label{fig:statsxD} Mean horizontal (a) and vertical (b) deflection of the wake, $\bar{y}_c$ and $\bar{z}_c$, at different downstream distances, for inflow 1, for every yaw and tilt case. For every case, the respective color is darker as $\gamma$ or $\beta$ increases. The red circles represent the FAST.Farm estimation, and the triangles joined by the blue lines, the VFS results. The $C_\text{meand}$ used for each case and downstream location is fitted depending on the downstream distance $x$, see Table~\ref{tab:cMeandX}. The dotted lines represent the estimation of the wake deflection in yawed conditions by the analytical model by Qian and Ishihara \cite{qian2018}.}
\end{figure}

The mean values for the horizontal and vertical deflection of the wake at $x = 2,~ 4,~ 6$ and $8D$ are shown in Figure~\ref{fig:statsxD}, for inflow 1. The blue circles and red triangle markers go from lighter to darker blue and red, respectively for the two models, as the deflection angles $\gamma$ or $\beta$ increase. The mean horizontal deflection at the different downstream locations is sensitive to the $C_\text{meand}$, as discussed by \cite{Cheng2018}. As the distance downstream decreases, the $C_\text{meand}$ used increases. For instance, for $\gamma = 10$\textdegree, at $x = 8D$, the filter $C_\text{meand}$ used is 1.90, whereas at $x = 6D$, $C_\text{meand}$ is 2.00. 
The $C_\text{meand}$ used for each case and location is fitted depending on the downstream distance $x$, for $x \in(2,8)D$. The fitted $C_\text{meand}$ filters used for each simulation to evaluate the horizontal mean deflection at each downstream distance are shown in Table~\ref{tab:cMeandX}. By using the filters adjusted for every downstream distance, for the cases $x \geq 2D$ the maximum difference observed is 7.4\%, for $x = 8D$ and $\gamma = 15$\textdegree. At $x = 2D$, the mean deflections values of FAST.Farm are overestimated, which is consistent with the fact that the near-wake modelling equations in OpenFAST are simplified with the far-wake model assumption. As the downstream distance increases, the mean values for FF compare better to those of VFS. The estimation for the mean horizontal deflection in this work is in agreement with the values estimated by the analytical model for the far-wake from Qian and Ishihara \cite{qian2018}, depicted by the dotted lines in Figure~\ref{fig:statsxD} a), for the different yaw angles. %The comparison to the analytical model by Bastankhah and Porté-Agel \cite{bastankhah2016} was not feasible, since the definition of some parameters in their model could not be defined for the current case. 

Regarding the mean vertical wake deflection, for $\beta = $10\textdegree, 20\textdegree\xspace and 30\textdegree\xspace, the values are close to the mean horizontal deflection for $\gamma = 10$\textdegree, 20\textdegree\xspace and 30\textdegree. The difference that is observed for the $\beta = 0$\textdegree\xspace and -6\textdegree\xspace cases is explained by how each model defines the wake center. First, in VFS, the instantaneous wake center is defined as the weight center of the velocity deficit, which is computed by subtracting the inflow velocity from the total velocity of the wake simulation. This approach removes the low velocity region near the ground due to the shear of the inflow, such that the low speed region of the inflow will not be mixed with the wake. However, such a definition results in a higher vertical position of the wake center than that defined using the streamwise velocity. Second, the wake center predicted by FAST.Farm is lower, since the ground effect is not modelled in this approach, but is implicitly imposed in VFS. The analysis for inflow 2 is very similar to that of inflow 1, and therefore is not presented. 

\begin{table}[h!]
\caption{ \label{tab:cMeandX} Fitted $C_\text{meand}$ filter values based on the results at $x \in (2,8D)$, for inflow 1. }
\centering
 \resizebox{0.6\linewidth}{!}{\begin{tabular}{c c | c c c c  }
Case$\quad\longrightarrow$ & 0  & 1a        & 2a       & 3a        & 4a                 \\
\hline
  $C_\text{meand}$ @ $x=2D$         & 2.67 & 2.10   & 2.20   & 2.00 & 2.30 \\
    $C_\text{meand}$ @ $x=4D$         & 2.67 & 2.10   & 2.20   & 2.20 & 2.30 \\
      $C_\text{meand}$ @ $x=6D$         & 2.67 &  2.00  &  2.10  & 2.10 & 2.20 \\
        $C_\text{meand}$ @ $x=8D$         & 2.67 & 1.90  & 2.00   & 2.00 & 2.10 \\
\hline
    \end{tabular}}
\end{table}

%{\color{green}Hi Zhaobin, could you please complete this with what we discussed on Friday, based on the wake definition in VFS}; furthermore, the ground effect is not modelled in FAST.Farm, whereas it is in VFS?. \color{black} 

%For the tilt deflection cases, the mean values for both TI levels inflow almost overlap. When comparing the mean values for the different yaw misalignment and tilt deflection angles, it is seen that the values are very similar, with a maximum difference of ...\% for the .... The value of the standard deviation yields a larger difference, especially for the negative $\beta$ angle. As the tilt angle increases this difference is decreased, up to....

\subsubsection{Time and frequency analysis of wake meandering based on the $C_\text{meand}$}\label{sec:Cmeand}

\begin{figure}[h!]
\centering
  \sbox0{\solidGreen} \sbox1{\solidRed} \sbox2{\solidBlue}\sbox3{\solidOrange}\sbox4{\solidPurple} 
    \framebox{\shortstack{ \usebox0 $C_\text{meand}=2.67$; \usebox1 $C_\text{meand}=1.90$, FAST.Farm; \usebox2 VFS\\
    \usebox3   $U$- \usebox4 $V$-component} }
\centering

\vspace{0.5cm}

\includegraphics[width=1\textwidth]{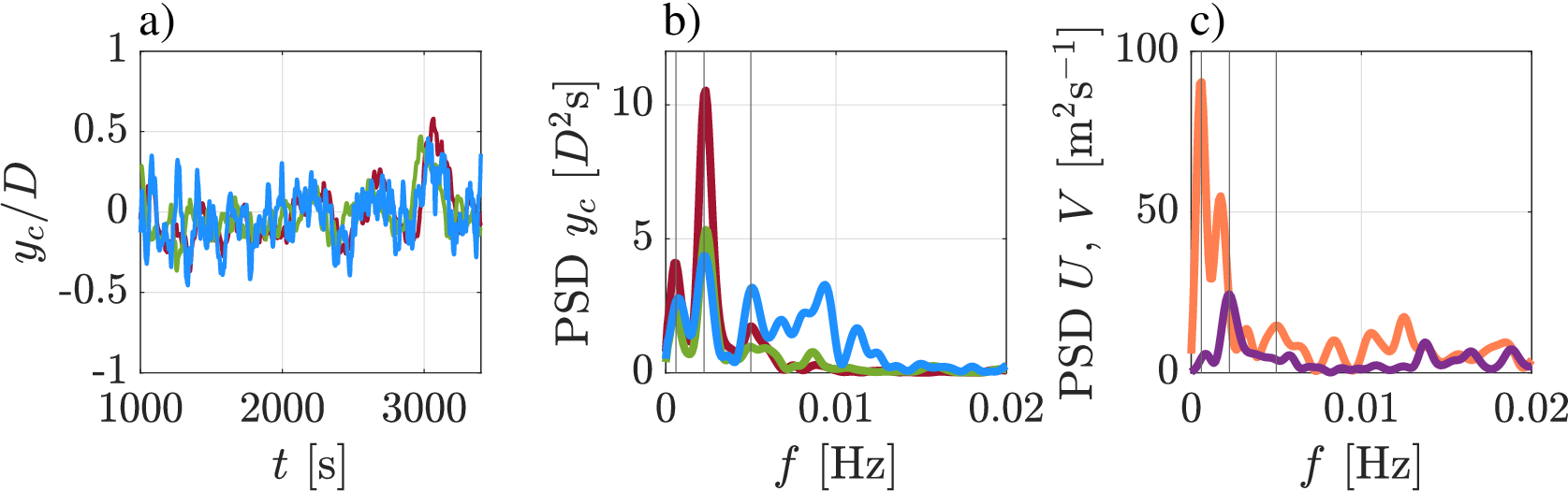}
\caption{ \label{fig:CMeand} Time series (a) and PSD (b) of the horizontal wake center meandering for $\gamma=0$\textdegree, for FAST.Farm and two $C_\text{meand}$ filters, and for VFS, at $x$= 8D, for inflow 1. PSD (c) of the incoming wind speed $U$ and $V$-components at the node at hub-height. The vertical lines indicate the frequencies f = 0.0006, 0.0022 and 0.005\,Hz. }
\end{figure}

To illustrate the frequency of wake meandering, the time series and power spectral density of the horizontal meandering $y_c$ for case 0, i.e. no horizontal or vertical wake deflection, is presented in Figure~\ref{fig:CMeand}, for two $C_\text{meand}$ filters in FAST.Farm. The $C_\text{meand}=1.90$, in red, and with a higher frequency content at the lower frequencies, is the approximate mean value of the filter used for cases 1a to 4a, i.e. the yaw misalignment cases; the $C_\text{meand}=2.67$ is the one based on the work by Cheng and Porté-Agel \cite{Cheng2018}. From both the time series (a) and the PSD (b), three main low frequency components at 0.0006 and 0.0022 and 0.0050\,Hz, outlined by vertical lines in the PSDs, are captured by the two models, and the different levels of filtering. 

Comparing the results in FAST.Farm depending on the $C_\text{meand}$, the smaller filter ($C_\text{meand}=1.90$), in red, presents a larger standard deviation, and is therefore closer to the VFS model. On the other hand, the higher energy content yielding this higher standard deviation is not at the same frequencies as the VFS model, i.e. at $f \approx 0.01$\,Hz. Instead, the energy content at 0.0006 and 0.0022 and 0.005\,Hz is higher for this smaller filter. These peaks are also observed in the $U,V$-components of the undisturbed incoming wind field, Figure~\ref{fig:PSDWF} and Figure~\ref{fig:CMeand}, c). Therefore, the main conclusion is that, regardless of the $C_\text{meand}$ filter used, the lower frequencies, related to the $U,V$-components of the incoming wind field, are, not only dominant in the meandering, but properly reproduced by FAST.Farm. However, higher frequency components, at approximately 0.01\,Hz, are not captured by the DWM as implemented in FAST.Farm. This higher energy content identified in the VFS case may be related to the shear-instability wake meandering mechanism, which has been characterized to fall in the range of a Strouhal number, defined as $St = fD/U_{\infty,z_{HH}}$, between 0.1 and 0.5 \cite{heisel2018}. In this case, $St$ is equal to 0.27, and based on qualitative analyses of videos of the flow field, the hypothesis of this higher energy content being related to this mechanism is plausible.

\begin{figure}[h!]
\centering
  \sbox0{\solidGreen} \sbox1{\solidRed} \sbox2{\solidBlue}\sbox3{\solidOrange}\sbox4{\solidYellow} 
    \framebox{\shortstack{ \usebox0 $C_\text{meand}=2.67$; \usebox1 $C_\text{meand}=2.90$, FAST.Farm; \usebox2 VFS\\
    \usebox3   $U$- \usebox4 $W$-component} }
\centering

\vspace{0.5cm}

\includegraphics[width=1\textwidth]{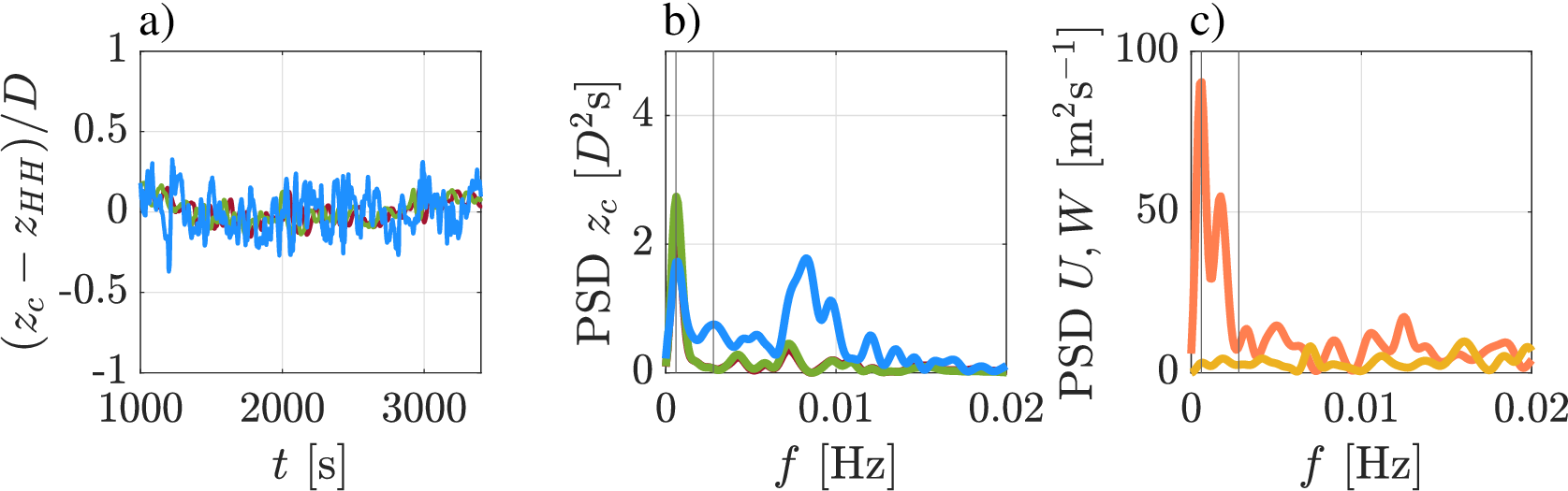}
\caption{ \label{fig:CMeandVert} Time series (a) and PSD (b) of the vertical wake center meandering for $\gamma=0$\textdegree, for FAST.Farm and two $C_\text{meand}$ filters, and for VFS, at $x= 8D$, for inflow 1. PSD (c) of the incoming wind speed $U$ and $W$-components at the node at hub-height $z_{HH}$. The vertical lines outline the frequencies $f$ = 0.0006 and 0.0028\,Hz. }
\end{figure}

Figure \ref{fig:CMeandVert} further investigates the vertical wake meandering for the case with $\gamma=0$ and $\beta = 0$. In general, the time history of the wake center vertical location  $z_c (t)$ predicted by both models is smaller compared to the meandering in the horizontal direction. From the DWM model point of view, the weaker wake meandering in the vertical direction is due to the inflow. The velocity fluctuation in the vertical direction is smaller than in the lateral direction, as shown in Table~\ref{tab:inflow}. For the LES, the ground also has a limiting effect to the vertical movement of the wake. Again, the vertical wake movement predicted by VFS shows a distinct frequency signature at $f \approx 0.01\,$Hz, presumably corresponding to a shear-layer-induced phenomenon, which is not captured by the DWM model as implemented in FAST.Farm. 

\begin{figure}[h!]
\centering
   \sbox0{\solidRed} \sbox1{\solidBlue}
    \framebox{\shortstack{\usebox0  FAST.Farm; \usebox1 VFS-Wind} }
        
 \vspace{0.2cm}   
\centering
\includegraphics[trim={0 0cm 0 0cm},clip,width=1\textwidth]{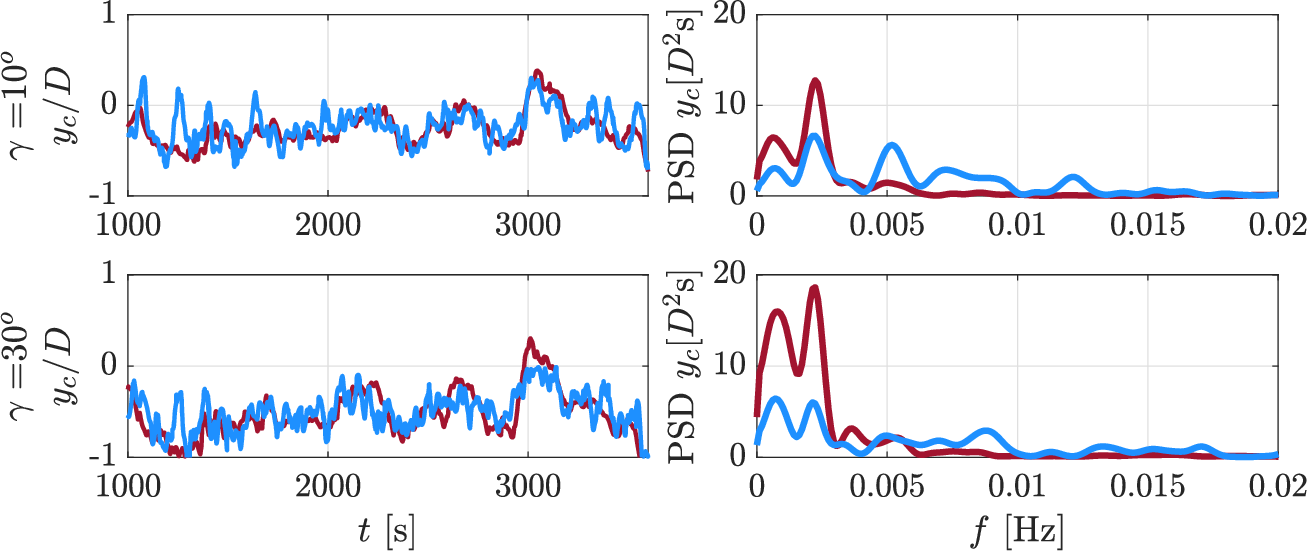}

\caption{ \label{fig:MS_meandTD} Time series and PSD of the horizontal wake center meandering $y_c$ at $x = 8D$, for the yaw misalignment cases with $\gamma = 10$\textdegree\xspace and 30\textdegree, for inflow 1. }
\end{figure}

Figure~\ref{fig:MS_meandTD} shows the time series of the horizontal meandering of the wake center $y_c (t)$ and the PSD for $\gamma = 10$\textdegree\xspace and $\gamma = 30$\textdegree, at $x = 8D$. The $C_\text{meand}$ used in these cases are the ones corresponding to Table~\ref{tab:cMeand}, respectively to the two yaw angle deflections. The trend is similar to the one observed in Figure~\ref{fig:CMeand}: the three peaks at the lower frequencies are captured, but the level of energy related to them is overestimated by FAST.Farm for the two lowest frequencies at 0.0006 and 0.0022\,Hz, and underestimated as the frequency increases. The overestimation is more visible for the larger deflection angle $\gamma = 30^{\circ}$. Furthermore, FAST.Farm doesn't capture the same trend at all with respect to the effect of yaw on the frequency content of the wake center position compared to VFS; the former increases the low-frequency energy content as the yaw angle increases, whereas for VFS this energy content decreases.\\

\begin{figure}[h!]
\centering
   \sbox0{\solidRed} \sbox1{\solidBlue}
    \framebox{\shortstack{\usebox0  FAST.Farm; \usebox1 VFS-Wind} }
    
 \vspace{0.2cm}   
\centering
\includegraphics[trim={0 0cm 0 0cm},clip,width=1\textwidth]{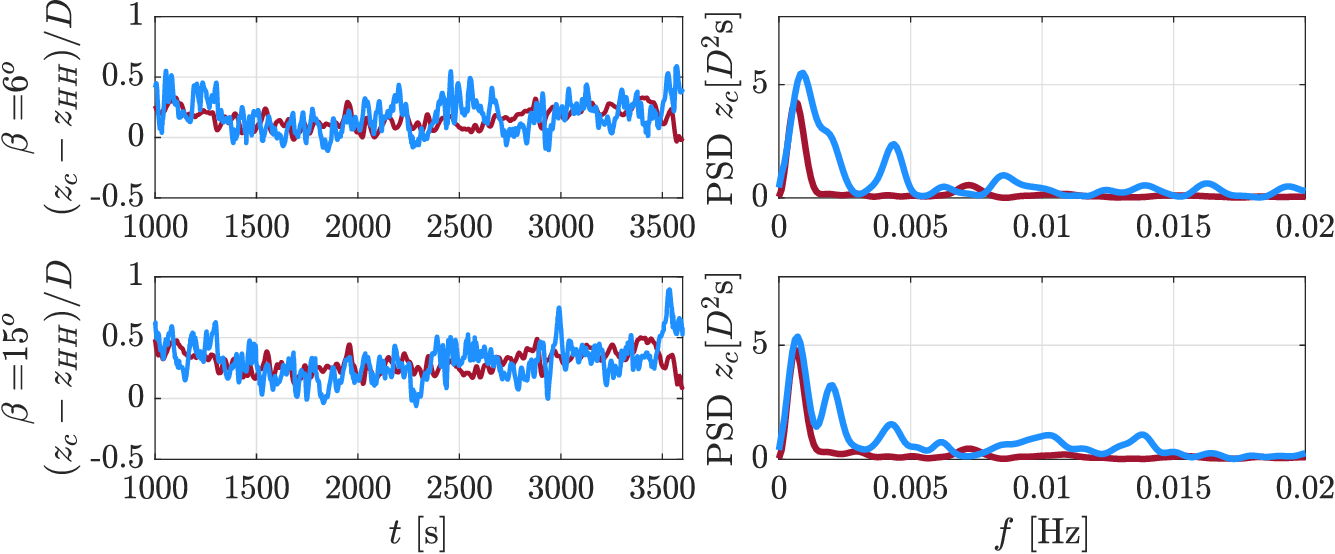}

\caption{ \label{fig:MS_meandTDT} Time series and PSD of the vertical wake center meandering $z_c$ at $x = 8D$, for the tilt misalignment angles $\beta = 6$\textdegree\xspace and 15\textdegree, for inflow 1. }
\end{figure}

The vertical meandering for $\beta=6$\textdegree\xspace and 15\textdegree, depicted in Figure~\ref{fig:MS_meandTDT}, shows that FAST.Farm captures the lowest frequency with a higher energy content at 0.0006\,Hz, for both tilt angle deflection cases, related to the incoming wind field. The higher frequency components, related to shear-layer instability meandering mechanisms, are not properly reproduced. This difference in the energy content for both tilt deflections is reflected in the standard deviation in Figure~\ref{fig:IEAComparisonMeand}, d), as a general trend, and specifically for $\beta=6$\textdegree\xspace and 15\textdegree. In FAST.Farm, the energy content for both the tilt deflection angles presented here is similar for the lowest frequency at 0.0006\,Hz. However, in VFS, the energy content decreases for this frequency whereas it increases for higher frequencies, as the tilt angle increases from 6\textdegree\xspace to 15\textdegree.

\newpage
\subsection{Effect of the yaw and tilt wake steering on the power output of a wind turbine downstream}

\begin{figure}[h!]
\centering
\sbox0{\solidPurple}\sbox1{\solidYellow}%
    \centering
    \framebox{\shortstack{\usebox0   inflow 1  , \usebox1 inflow 2 } }
\centering

\vspace{0.5cm}

\includegraphics[width=1\textwidth]{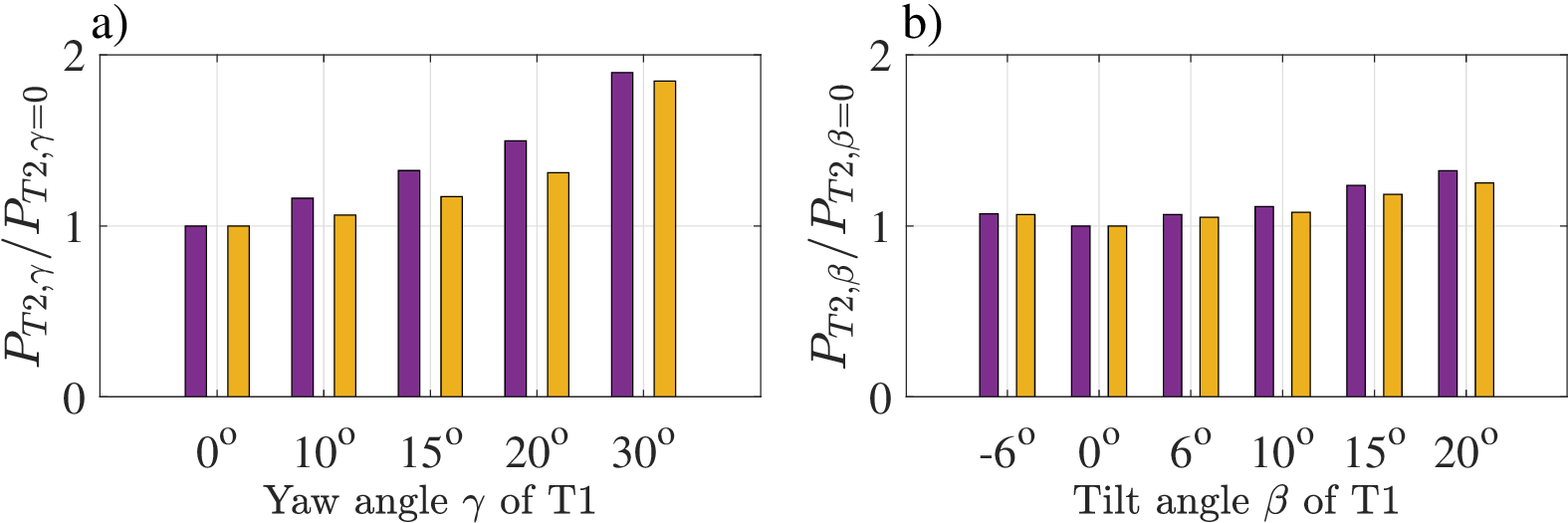}
\caption{ \label{fig:pOut} Power output of the turbine in the wake $P_{T2,\gamma}$ (a) and $P_{T2,\beta}$ (b), for every case of yaw and tilt misalignment, respectively, and for the two turbulent inflows, inflow 1 and inflow 2. The power output is normalized by the power output of the same turbine (T2) if $\beta$ and $\gamma$ of the upstream turbine are zero ($P_{T2,\beta,\gamma = 0}$).}
\end{figure}

To estimate how the deflection of the wake may affect the power output of a wind turbine in the wake, the same set of cases was run in FAST.Farm, with a second IEA 15MW wind turbine (T2) placed $8D$ downstream the first turbine in free wind (T1). T2 has an operating controller, i.e. not simplified to match the LES rotor configuration. Figure~\ref{fig:pOut} shows the increase in power output of T2 as the wake deflects due to the yaw misalignment or rotor tilt of the wind turbine in free wind. The results are normalized by the power output of T2 if T1 were not yawed or tilted ($P_{T2,\gamma=0}$ and $P_{T2,\beta=0}$, respectively). The power output of the turbine in the wake for $\gamma = 30$\textdegree\xspace is almost doubled compared to completely waked conditions. 

\section{Conclusions}\label{sec:conclusions}

To reduce the negative effects of the wakes on the downstream wind turbines in a wind farm, several turbine control strategies have been developed and evaluated. One of them is wake steering, which changes the direction of wakes by setting deliberately a misalignment between the rotor and the wind. Possible misalignment strategies include yaw-based and tilt-based controls \cite{fleming2014,aitken2014a}, which deflect the wake in the lateral and horizontal directions, respectively.

In this work, twenty yaw ($\gamma$) and tilt ($\beta$) wind turbine misalignment cases are studied for the IEA 15MW wind turbine. Two inflows with different shear and turbulence intensities and a mean wind speed of approximately 9\,m/s are investigated. For this purpose, large eddy simulations (VFS-Wind) and the dynamic wake meandering model (FAST.Farm) are used. The focus is on the time-averaged velocity field and wake deficit downstream the wind turbine, the average wake deflection, vertical and horizontal, and the meandering, or standard deviation of the wake center displacement, at 8 rotor diameters ($D$) downstream. The main conclusions on the validation of the results by FAST.Farm, based on the high-fidelity results by VFS-Wind, include:

\begin{itemize}
 \item The horizontal wake steering, including the time-averaged wake deflection and velocity profiles show very good agreement for all the cases presented, especially for $x \geq 6D$. The vertical wake steering shows very good agreement for cases with positive tilt angles ($\beta>6^\circ$). For $\beta \le 0 ^\circ$, discrepancies are observed when analysing the mean value, mainly due to whether or not the presence of the ground is accounted for.

 \item The mean and standard deviation of the horizontal and vertical meandering computed by the DWM (FAST.Farm) are sensitive to the size of the polar grid used to calculate the spatial-averaged velocity with which the wake planes are meandered. This size is defined by the $C_\text{meand}$ parameter in FAST.Farm. As this filter increases, so does the polar grid area used to calculate this spatial-averaged velocity, and the lower impact the deficit has on the averaged longitudinal component. This lower impact of the deficit leads to a smaller ($V,W$)-to-$U$ ratio, and therefore to lower mean and standard deviation of the wake center meandering.  

    \item Both the DWM and LES approaches capture the large eddy-induced wake meandering at $St < 0.1$. In addition, LES captures the wake oscillation induced by the wake shear layer at $St \approx 0.27$.

\end{itemize}

FAST.Farm shows a generally good performance for the cases presented here. The main limitations of this tool as used in this work which affect the wake deficit and meandering are listed in the following. First, ground effects are not modelled, which has an impact mainly on the vertical wake steering. Second, the wake meandering frequencies related to shear-layer instability are not captured; these effects are observed at approximately 0.01\,Hz, which coincides with the typical natural periods of floating wind turbines. The last main limitation, based on the findings in this work, is related to the $C_\text{meand}$: this filter cannot be adjusted for the vertical and horizontal deflection separately, which entails the limitation of not being able to, for instance, analyse a yaw-control strategy in floating wind farms, since fitting the $C_\text{meand}$ for a specific yaw angle will yield an overestimation of the mean vertical wake deflection. Additionally, the current FAST.Farm DWM implementation considers a fixed $C_\text{meand}$ in time and space in a given simulation, while the current results suggest that a streamwise variation might be useful. \\ 

Despite the limitations of the mid-fidelity tool, FAST.Farms provides a good compromise between its efficiency and accuracy, given the reduced computational cost: in the current case, FAST.Farm requires 3 hour runtime to run 3600\,s simulation on a single core, whereas VFS-Wind uses 48 hours on 240 cores. Nevertheless, the power output gain of a wind turbine in the wake of the yawed and tilted rotors analysed in this work could be computed by FAST.Farm. From this analysis it was concluded that if the wind turbine in free-wind is yawed by $\gamma = 30^{\circ}$, the power output of a turbine placed 8$D$ downstream is almost doubled compared to if the turbine in free wind were not yawed.\\

Due to the computational cost constraints of LES, running the cases shown in this work to analyse the wake steering for other mean wind speed scenarios, and for more than one seed, was not feasible. However, future work should consider additional cases, and perform a similar analysis using the curled wake formulation implemented FAST.Farm.

%. It is an efficient method to estimate of wake and inter-array phenomena effects on wind turbines in utility-scale wind farms
\begin{comment}

\end{comment}
%TC:ignore
\section*{Acknowledgements}

The research leading to these results has received funding from the Basic Science Center Program for ``Multiscale Problems in Nonlinear Mechanics'' of the National Natural Science Foundation of China (NO. 11988102), and by the Research Council of Norway through the ENERGIX programme (grant 294573) and industry partners Equinor, MacGregor, Inocean, APL Norway and RWE Renewables, and NSFC (NO. 12172360, NO.12202453). 
%TC:endignore
\clearpage

 \bibliographystyle{bibstyle} 
 \bibliography{main}

\begin{thebibliography}{10}
\expandafter\ifx\csname url\endcsname\relax
  \def\url#1{\texttt{#1}}\fi
\expandafter\ifx\csname urlprefix\endcsname\relax\def\urlprefix{URL }\fi
\expandafter\ifx\csname href\endcsname\relax
  \def\href#1#2{#2} \def\path#1{#1}\fi

\bibitem{barthelmie2010}
R.~J. Barthelmie, L.~E. Jensen, {Evaluation of wind farm efficiency and wind
  turbine wakes at the Nysted offshore wind farm}, Wind Energy 13~(6) (2010)
  573--586.
\newblock \href {https://doi.org/10.1002/we.408} {\path{doi:10.1002/we.408}}.

\bibitem{stevens2017}
R.~J. A.~M. Stevens, C.~Meneveau, Flow structure and turbulence in wind farms,
  Annual Review of Fluid Mechanics 49~(1) (2017) 311--339.
\newblock \href {https://doi.org/10.1146/annurev-fluid-010816-060206}
  {\path{doi:10.1146/annurev-fluid-010816-060206}}.

\bibitem{simley2020}
E.~Simley, P.~Fleming, J.~King, Field validation of wake steering control with
  wind direction variability, Journal of Physics: Conference Series 1452~(1)
  (2020) 012012.
\newblock \href {https://doi.org/10.1088/1742-6596/1452/1/012012}
  {\path{doi:10.1088/1742-6596/1452/1/012012}}.

\bibitem{wise2019}
A.~S. Wise, E.~E. Bachynski, Wake meandering effects on floating wind turbines,
  {Wind Energy} 23~(5) (2019).
\newblock \href {https://doi.org/10.1002/we.2485} {\path{doi:10.1002/we.2485}}.

\bibitem{nanos2021}
E.~M. Nanos, C.~L. Bottasso, D.~I. Manolas, V.~A. Riziotis, Vertical wake
  deflection for floating wind turbines by differential ballast control, {Wind
  Energy Science} (aug 2021).
\newblock \href {https://doi.org/10.5194/wes-2021-79}
  {\path{doi:10.5194/wes-2021-79}}.

\bibitem{doubrawa2021}
P.~Doubrawa, S.~Sirnivas, M.~Godvik, Effects of upstream rotor tilt on a
  downstream floating wind turbine, Journal of Physics: Conference Series
  1934~(1) (2021) 012016.
\newblock \href {https://doi.org/10.1088/1742-6596/1934/1/012016}
  {\path{doi:10.1088/1742-6596/1934/1/012016}}.

\bibitem{bastankhah2016}
M.~Bastankhah, F.~Port{\'{e}}-Agel, Experimental and theoretical study of
  wind~turbine wakes in yawed conditions, Journal of Fluid Mechanics 806 (2016)
  506--541.
\newblock \href {https://doi.org/10.1017/jfm.2016.595}
  {\path{doi:10.1017/jfm.2016.595}}.

\bibitem{fleming2014}
P.~Fleming, P.~M. Gebraad, S.~Lee, J.-W. van Wingerden, K.~Johnson,
  M.~Churchfield, J.~Michalakes, P.~Spalart, P.~Moriarty, Simulation comparison
  of wake mitigation control strategies for a two-turbine case, Wind Energy
  18~(12) (2014) 2135--2143.
\newblock \href {https://doi.org/10.1002/we.1810} {\path{doi:10.1002/we.1810}}.

\bibitem{howland2016}
M.~F. Howland, J.~Bossuyt, L.~A. Mart{\'{\i}}nez-Tossas, J.~Meyers,
  C.~Meneveau, Wake structure in actuator disk models of wind turbines in yaw
  under uniform inflow conditions, Journal of Renewable and Sustainable Energy
  8~(4) (2016) 043301.
\newblock \href {https://doi.org/10.1063/1.4955091}
  {\path{doi:10.1063/1.4955091}}.

\bibitem{archer2019}
C.~L. Archer, A.~Vasel-Be-Hagh, Wake steering via yaw control in multi-turbine
  wind farms: recommendations based on large-eddy simulation, Sustainable
  Energy Technologies and Assessments 33 (2019) 34--43.
\newblock \href {https://doi.org/10.1016/j.seta.2019.03.002}
  {\path{doi:10.1016/j.seta.2019.03.002}}.

\bibitem{aitken2014a}
M.~L. Aitken, R.~M. Banta, Y.~L. Pichugina, J.~K. Lundquist, Quantifying wind
  turbine wake characteristics from scanning remote sensor data, Journal of
  Atmospheric and Oceanic Technology 31~(4) (2014) 765--787.
\newblock \href {https://doi.org/10.1175/jtech-d-13-00104.1}
  {\path{doi:10.1175/jtech-d-13-00104.1}}.

\bibitem{qian2018}
G.-W. Qian, T.~Ishihara, A new analytical wake model for yawed wind turbines,
  Energies 11~(3) (2018) 665.
\newblock \href {https://doi.org/10.3390/en11030665}
  {\path{doi:10.3390/en11030665}}.

\bibitem{howland2022collective}
M.~F. Howland, J.~B. Quesada, J.~J.~P. Mart{\'\i}nez, F.~P. Larra{\~n}aga,
  N.~Yadav, J.~S. Chawla, V.~Sivaram, J.~O. Dabiri, Collective wind farm
  operation based on a predictive model increases utility-scale energy
  production, Nature Energy 7~(9) (2022) 818--827.

\bibitem{fleming2014a}
P.~A. Fleming, P.~M. Gebraad, S.~Lee, J.-W. van Wingerden, K.~Johnson,
  M.~Churchfield, J.~Michalakes, P.~Spalart, P.~Moriarty, Evaluating techniques
  for redirecting turbine wakes using {SOWFA}, Renewable Energy 70 (2014)
  211--218.
\newblock \href {https://doi.org/10.1016/j.renene.2014.02.015}
  {\path{doi:10.1016/j.renene.2014.02.015}}.

\bibitem{annoni2017}
J.~Annoni, A.~Scholbrock, M.~Churchfield, P.~Fleming, Evaluating tilt for wind
  plants, in: 2017 American Control Conference ({ACC}), {IEEE}, 2017, pp.
  1--20.
\newblock \href {https://doi.org/10.23919/acc.2017.7963037}
  {\path{doi:10.23919/acc.2017.7963037}}.

\bibitem{johlas2022}
H.~M. Johlas, D.~P. Schmidt, M.~A. Lackner, Large eddy simulations of curled
  wakes from tilted wind turbines, Renewable Energy 188 (2022) 349--360.
\newblock \href {https://doi.org/10.1016/j.renene.2022.02.018}
  {\path{doi:10.1016/j.renene.2022.02.018}}.

\bibitem{cossu2020}
C.~Cossu, Replacing wakes with streaks in wind turbine arrays, Wind Energy
  24~(4) (2020) 345--356.
\newblock \href {https://doi.org/10.1002/we.2577} {\path{doi:10.1002/we.2577}}.

\bibitem{cossu2021}
C.~Cossu, Evaluation of tilt control for wind-turbine arrays in the atmospheric
  boundary layer, Wind Energy Science 6~(3) (2021) 663--675.
\newblock \href {https://doi.org/10.5194/wes-6-663-2021}
  {\path{doi:10.5194/wes-6-663-2021}}.

\bibitem{larsen2008}
G.~C. Larsen, H.~A. Madsen, K.~Thomsen, T.~J. Larsen, Wake meandering: a
  pragmatic approach, {Wind Energy} 11~(4) (2008) 377--395.
\newblock \href {https://doi.org/10.1002/we.267} {\path{doi:10.1002/we.267}}.

\bibitem{churchfield2015}
M.~J. Churchfield, S.~Lee, P.~J. Moriarty, Y.~Hao, M.~A. Lackner,
  R.~Barthelmie, J.~K. Lundquist, G.~Oxley, {{A comparison of the Dynamic Wake
  Meandering model, Large-Eddy Simulation, and field data at the Egmond aan Zee
  offshore wind plant}}, in: 33rd Wind Energy Symposium, American Institute of
  Aeronautics and Astronautics, 2015, pp. 1--20.
\newblock \href {https://doi.org/10.2514/6.2015-0724}
  {\path{doi:10.2514/6.2015-0724}}.

\bibitem{jonkman2018validation}
J.~Jonkman, P.~Doubrawa, N.~Hamilton, J.~Annoni, P.~Fleming, {Validation of
  {FAST}.Farm against Large-Eddy Simulations}, Journal of Physics: Conference
  Series 1037 (2018) 062005.
\newblock \href {https://doi.org/10.1088/1742-6596/1037/6/062005}
  {\path{doi:10.1088/1742-6596/1037/6/062005}}.

\bibitem{Gaertner2020}
E.~Gaertner, J.~Rinker, L.~Sethuraman, F.~Zahle, B.~Anderson, G.~Barter,
  N.~Abbas, F.~Meng, P.~Bortolotti, W.~Skrzypinski, G.~Scott, R.~Feil,
  H.~Bredmose, K.~Dykes, M.~Shields, C.~Allen, A.~Viselli, {IEA} wind {TCP}
  task 37: Definition of the {IEA} 15-megawatt offshore reference wind turbine,
  Tech. rep., NREL (mar 2020).
\newblock \href {https://doi.org/10.2172/1603478} {\path{doi:10.2172/1603478}}.

\bibitem{yang2015VFS}
X.~Yang, F.~Sotiropoulos, R.~J. Conzemius, J.~N. Wachtler, M.~B. Strong,
  Large‐eddy simulation of turbulent flow past wind turbines/farms: the
  virtual wind simulator ({VWiS}), Wind Energy 18~(12) (2015) 2025--2045.

\bibitem{Smagorinsky}
J.~Smagorinsky, General circulation experiments with the primitive equations:
  I. the basic experiment, Monthly Weather Review 91~(3) (1963) 99--164.

\bibitem{yang2018ASMethod}
X.~Yang, F.~Sotiropoulos, A new class of actuator surface models for wind
  turbines, Wind Energy 21~(5) (2018) 285--302.

\bibitem{knoll2004jacobian}
D.~A. Knoll, D.~E. Keyes, Jacobian-free {Newton--Krylov} methods: a survey of
  approaches and applications, Journal of Computational Physics 193~(2) (2004)
  357--397.

\bibitem{saad1993flexible}
Y.~Saad, A flexible inner-outer preconditioned {GMRES} algorithm, SIAM Journal
  on Scientific Computing 14~(2) (1993) 461--469.

\bibitem{ge2007numerical}
L.~Ge, F.~Sotiropoulos, A numerical method for solving the {3D} unsteady
  incompressible {Navier--Stokes} equations in curvilinear domains with complex
  immersed boundaries, {Journal of Computational Physics} 225~(2) (2007)
  1782--1809.

\bibitem{yang2016coherent}
X.~Yang, J.~Hong, M.~Barone, F.~Sotiropoulos, Coherent dynamics in the rotor
  tip shear layer of utility-scale wind turbines, Journal of Fluid Mechanics
  804 (2016) 90--115.

\bibitem{hong2020snow}
J.~Hong, A.~Abraham, Snow-powered research on utility-scale wind turbine flows,
  Acta Mechanica Sinica 36 (2020) 339--355.

\bibitem{foti2021coherent}
D.~Foti, Coherent vorticity dynamics and dissipation in a utility-scale wind
  turbine wake with uniform inflow, Theoretical and Applied Mechanics Letters
  11~(5) (2021) 100292.

\bibitem{li2021}
Z.~Li, X.~Yang, Large-eddy simulation on the similarity between wakes of wind
  turbines with different yaw angles, Journal of Fluid Mechanics 921 (jun
  2021).
\newblock \href {https://doi.org/10.1017/jfm.2021.495}
  {\path{doi:10.1017/jfm.2021.495}}.

\bibitem{li2022onset}
Z.~Li, G.~Dong, X.~Yang, Onset of wake meandering for a floating offshore wind
  turbine under side-to-side motion, Journal of Fluid Mechanics 934 (2022) A29.

\bibitem{yang2018large}
X.~Yang, M.~Pakula, F.~Sotiropoulos, Large-eddy simulation of a utility-scale
  wind farm in complex terrain, Applied energy 229 (2018) 767--777.

\bibitem{wang2023statistics}
Z.~Wang, G.~Dong, Z.~Li, X.~Yang, Statistics of wind farm wakes for different
  layouts and ground roughness, Boundary-Layer Meteorology (2023) 1--36.

\bibitem{jonkman2021}
J.~Jonkman, K.~Shaler, FAST.Farm User’s Guide and Theory Manual, NREL,
  Boulder, Colorado (2021).

\bibitem{OF34}
{OpenFAST 3.4.0}, \url{hhttps://github.com/OpenFAST/openfast} (2023).

\bibitem{madsen2010}
H.~A. Madsen, G.~C. Larsen, T.~J. Larsen, N.~Troldborg, R.~Mikkelsen,
  {{Calibration and validation of the Dynamic Wake Meandering model for
  implementation in an aeroelastic code}}, Journal of Solar Energy Engineering
  132~(4) (oct 2010).
\newblock \href {https://doi.org/10.1115/1.4002555}
  {\path{doi:10.1115/1.4002555}}.

\bibitem{keck2013PhD}
R.~E. Keck, A consistent turbulence formulation for the dynamic wake meandering
  model in the atmospheric boundary layer, Ph.D. thesis, DTU, Denmark (2013).

\bibitem{doubrawa2018}
P.~Doubrawa, J.~R. Annoni, J.~M. Jonkman, {Optimization-based calibration of
  {FAST}.Farm parameters against Large-Eddy Simulations}, in: 2018 Wind Energy
  Symposium, American Institute of Aeronautics and Astronautics, 2018, pp.
  1--20.
\newblock \href {https://doi.org/10.2514/6.2018-0512}
  {\path{doi:10.2514/6.2018-0512}}.

\bibitem{branlard2022}
E.~Branlard, L.~A. Mart{\'{\i}}nez-Tossas, J.~Jonkman, A time-varying
  formulation of the curled wake model within the {FAST}.farm framework, Wind
  Energy (sep 2022).
\newblock \href {https://doi.org/10.1002/we.2785} {\path{doi:10.1002/we.2785}}.

\bibitem{Oye1991}
S.~Øye, Dynamic stall simulated as time lag of separation, Proceedings of the
  Fourth IEA Symposium on the Aerodynamics of Wind Turbines (nov 1990).

\bibitem{Cheng2018}
W.-C. Cheng, F.~Port{\'{e}}-Agel, A simple physically-based model for
  wind-turbine wake growth in a turbulent boundary layer, Boundary-Layer
  Meteorology 169~(1) (2018) 1--10.
\newblock \href {https://doi.org/10.1007/s10546-018-0366-2}
  {\path{doi:10.1007/s10546-018-0366-2}}.

\bibitem{Brugger2022}
P.~Brugger, C.~Markfort, F.~Port{\'{e}}-Agel, Field measurements of wake
  meandering at a utility-scale wind turbine with nacelle-mounted doppler
  lidars, Wind Energy Science 7~(1) (2022) 185--199.
\newblock \href {https://doi.org/10.5194/wes-7-185-2022}
  {\path{doi:10.5194/wes-7-185-2022}}.

\bibitem{Yang2019}
X.~Yang, F.~Sotiropoulos, Wake characteristics of a utility-scale wind turbine
  under coherent inflow structures and different operating conditions, Phys.
  Rev. Fluids 4 (2019) 024604.
\newblock \href {https://doi.org/10.1103/PhysRevFluids.4.024604}
  {\path{doi:10.1103/PhysRevFluids.4.024604}}.

\bibitem{du19983Dstall}
Z.~Du, M.~Selig, A {3-D} stall-delay model for horizontal axis wind turbine
  performance prediction, in: 1998 ASME Wind Energy Symposium, 1998, p.~21.

\bibitem{shen2005tiploss2}
W.~Z. Shen, R.~Mikkelsen, J.~N. S{\o}rensen, C.~Bak, Tip loss corrections for
  wind turbine computations, Wind Energy 8~(4) (2005) 457--475.

\bibitem{Glauert1935}
H.~Glauert, Airplane propellers, in: Aerodynamic Theory, Springer Berlin
  Heidelberg, 1935, pp. 169--360.
\newblock \href {https://doi.org/10.1007/978-3-642-91487-43}
  {\path{doi:10.1007/978-3-642-91487-43}}.

\bibitem{Burton2011}
T.~Burton, N.~Jenkins, D.~Sharpe, E.~Bossanyi, Wind Energy Handbook, Wiley,
  2011.
\newblock \href {https://doi.org/10.1002/9781119992714}
  {\path{doi:10.1002/9781119992714}}.

\bibitem{Krogstad2011}
P.-{\AA}. Krogstad, M.~S. Adaramola, Performance and near wake measurements of
  a model horizontal axis wind turbine, Wind Energy 15~(5) (2011) 743--756.
\newblock \href {https://doi.org/10.1002/we.502} {\path{doi:10.1002/we.502}}.

\bibitem{Bartl2018}
J.~Bartl, Experimental testing of wind turbine wake control methods, Ph.D.
  thesis, NTNU (2018).

\bibitem{wang2016very}
G.~Wang, X.~Zheng, Very large scale motions in the atmospheric surface layer: a
  field investigation, Journal of Fluid Mechanics 802 (2016) 464--489.

\bibitem{liu2019three}
H.~Liu, G.~Wang, X.~Zheng, Three-dimensional representation of large-scale
  structures based on observations in atmospheric surface layers, Journal of
  Geophysical Research: Atmospheres 124~(20) (2019) 10753--10771.

\bibitem{blocken_cfd_2007}
B.~Blocken, T.~Stathopoulos, J.~Carmeliet, {CFD} simulation of the atmospheric
  boundary layer: wall function problems, Atmospheric Environment 41~(2) (2007)
  238 -- 252.

\bibitem{irwin1979}
J.~S. Irwin, A theoretical variation of the wind profile power-law exponent as
  a function of surface roughness and stability, Atmospheric Environment (1967)
  13~(1) (1979) 191--194.
\newblock \href {https://doi.org/10.1016/0004-6981(79)90260-9}
  {\path{doi:10.1016/0004-6981(79)90260-9}}.

\bibitem{emeis2014}
S.~Emeis, Current issues in wind energy meteorology, {Meteorological
  Applications} 21~(4) (2014) 803--819.
\newblock \href {https://doi.org/10.1002/met.1472}
  {\path{doi:10.1002/met.1472}}.

\bibitem{IEC61400-1}
IEC, {IEC61400-1: Wind energy generation systems– Part 1: Design
  requirements}, Tech. rep., IEC, Geneva, Switzerland (2019).

\bibitem{shaler2019}
K.~Shaler, J.~Jonkman, N.~Hamilton, {Effects of inflow spatiotemporal
  discretization on wake meandering and turbine structural response using
  {FAST}.Farm}, Journal of Physics: Conference Series 1256 (2019) 012023.
\newblock \href {https://doi.org/10.1088/1742-6596/1256/1/012023}
  {\path{doi:10.1088/1742-6596/1256/1/012023}}.

\bibitem{porte2020wind}
F.~Port{\'e}-Agel, M.~Bastankhah, S.~Shamsoddin, Wind-turbine and wind-farm
  flows: A review, Boundary-layer meteorology 174~(1) (2020) 1--59.

\bibitem{burton2011wind}
T.~Burton, N.~Jenkins, D.~Sharpe, E.~Bossanyi, Wind energy handbook, John Wiley
  \& Sons, 2011.

\bibitem{howland2016wake}
M.~F. Howland, J.~Bossuyt, L.~A. Mart{\'\i}nez-Tossas, J.~Meyers, C.~Meneveau,
  Wake structure in actuator disk models of wind turbines in yaw under uniform
  inflow conditions, Journal of Renewable and Sustainable Energy 8~(4) (2016).

\bibitem{heisel2018}
M.~Heisel, J.~Hong, M.~Guala, The spectral signature of wind turbine wake
  meandering: A wind tunnel and field-scale study, Wind Energy 21~(9) (2018)
  715--731.
\newblock \href {https://doi.org/10.1002/we.2189} {\path{doi:10.1002/we.2189}}.

\end{thebibliography}





% \end{thebibliography}
\end{document}